\documentclass[twocolumn,showpacs,amsmath,amssymb,floatfix,pre,floatfix]{revtex4} 

\usepackage{graphicx,color}
\definecolor{brown}{rgb}{0.63,0.27,0.18}
\definecolor{orange}{rgb}{1.00,0.65,0.00}

\usepackage{moresize}
\usepackage{dcolumn}
\usepackage{bm,multirow}

\begin{document}

\newcommand {\rsq}[1]{\langle R^2 (#1)\rangle}
\newcommand {\rsqL}{\langle R^2 (L) \rangle}
\newcommand {\rsqbp}{\langle R^2 (N_{bp}) \rangle}
\newcommand {\Nbp}{N_{bp}}
\newcommand {\etal}{{\em et al.}}
\newcommand{\Ham}{{\cal H}}
\newcommand{\RalfNew}[1]{\textcolor{red}{#1}}
\newcommand{\scs}{\ssmall}



\title{Computer simulations of randomly branching polymers: Annealed {\it vs.} quenched branching structures}

\author{Angelo Rosa}
\affiliation{SISSA - Scuola Internazionale Superiore di Studi Avanzati, Via Bonomea 265, 34136 Trieste, Italy}
\email{anrosa@sissa.it}

\author{Ralf Everaers}
\affiliation{Univ Lyon, Ens de Lyon, Univ Claude Bernard Lyon 1, CNRS, Laboratoire de Physique and Centre Blaise Pascal, F-69342 Lyon, France}
\email{ralf.everaers@ens-lyon.fr}

\date{\today}

\begin{abstract}
We present computer simulations of three systems of randomly branching polymers in $d=3$ dimensions:
ideal trees and self-avoiding trees with annealed and quenched connectivities. 
In all cases, we performed a detailed analysis of trees connectivities, spatial conformations
and statistical properties of linear paths on trees,
and compare the results to the corresponding predictions of Flory theory.
We confirm that, overall, the theory predicts correctly that trees with quenched ideal connectivity exhibit {\it less} overall swelling in good solvent than corresponding trees with annealed connectivity
even though they are {\it more strongly} stretched on the path level.
At the same time, we emphasize the inadequacy of the Flory theory in predicting the behaviour of other, and equally relevant, observables like contact probabilities between tree nodes.
We show, then, that contact probabilities can be aptly characterized by introducing a novel critical exponent, $\theta_{path}$, which accounts for how they decay as a function of the node-to-node path distance on the tree.
\end{abstract}

\maketitle

\section{Introduction}\label{sec:intro}
Randomly branched polymers or trees are of interest in a number of scientific fields.
Branched polymers can be synthesised by deliberately incorporating monomers with higher functionality into the polymerisation processes as a means of modifying materials properties~\cite{RubinsteinColby,Burchard1999}.
In industrial applications they represent the norm rather than the exception, since most polymerisation processes for linear chains also introduce a certain amount of branching, a feature that strongly affects the dynamics~\cite{Read2013}. 
In Statistical Mechanics, randomly branched polymers on a lattice are often referred to as lattice trees. They are believed to fall into the same universality class as lattice animals~\cite{IsaacsonLubensky,SeitzKlein1981,DuarteRuskin1981},
so that the critical exponents characterising them are related to those of magnetic systems~\cite{ParisiSourlasPRL1981,FisherPRL1978,KurtzeFisherPRB1979,BovierFroelichGlaus1984}.
Our own interest in these systems~\cite{GrosbergSoftMatter2014,RosaEveraersPRL2014,RubinsteinLeiden} is due to the analogy between their behavior and the crumpling of topologically constrained ring polymers~\cite{KhokhlovNechaev85,RubinsteinPRL1986,RubinsteinPRL1994} and, ultimately, chromosomes~\cite{grosbergEPL1993,RosaPLOS2008,Vettorel2009,MirnyRev2011}.

As customary in polymer physics~\cite{DeGennesBook,DoiEdwards,KhokhlovGrosberg,RubinsteinColby}, we are primarily interested in exponents describing how expectation values for observables scale with the weight, $N$, of the trees: 
\begin{eqnarray}
\langle N_{br}(N) \rangle & \sim & N^\epsilon                           \label{eq:epsilon}\\
\langle L(N) \rangle & \sim & N^\rho                                          \label{eq:rho}\\
\langle R_g^2(N) \rangle & \sim & N^{2\nu}                              \label{eq:nu}\\
\langle R^2(l) \rangle & \sim & l^{2\nu_{path}}                          \label{eq:nu_path}\\
\langle p_c(l) \rangle & \sim & l^{-\nu_{path}(d+\theta_{path})} \label{eq:theta_path}
\end{eqnarray}
Here, $\langle N_{br}(N)\rangle$ denotes the average branch weight; 
$\langle L(N) \rangle$ the average contour distance or length of paths on the tree; 
$\langle R_g^2(N) \rangle$ the mean-square gyration radius of the trees; 
and $\langle R^2(l) \rangle$ and $\langle p_c(l) \rangle$ the mean-square spatial distance and contact probability of nodes as a function of their contour distance, $l$.

The behaviour of trees depends on whether monomers interact or not, whether the tree connectivity is quenched or annealed, and whether the solutions are dilute or dense.
Typically, randomly branched polymers are considerably more compact than their linear counterparts.
For isolated, non-interacting trees $\nu=1/4$ compared to $\nu=1/2$ for linear chains~\cite{ZimmStockmayer49}. In $d=3$ the analogy to lattice animals allows to obtain the exact result $\nu=1/2$ for isolated, self-avoiding lattice trees with annealed connectivity \cite{ParisiSourlasPRL1981}.
In the polymer context, one has to distinguish the environmental conditions under which chains are synthesised, from those under which they are studied.
While the solvent quality in later experiments may vary, the chains always preserve a memory of the initial conditions under which their connectivity was ``quenched''.
Thus even if branched polymers can be described as having quenched ideal random connectivity, they fall into different universality class from polymers with annealed connectivity~\cite{GutinGrosberg93}
and will behave differently when studied in good solvent.
The peculiarity of the crumpled ring polymers is that the branched structure represents transient folding,
so that they map to lattice trees with {\em annealed} connectivity~\cite{KhokhlovNechaev85,RubinsteinPRL1986,RubinsteinPRL1994}. 

The exponents defined in Eqs.~(\ref{eq:epsilon}) to (\ref{eq:theta_path})
can be estimated within Flory theory~\cite{IsaacsonLubensky,DaoudJoanny1981,GutinGrosberg93,GrosbergSoftMatter2014}, which provides a simple, insightful and unifying description of all the different cases discussed above.
However, as for all Flory theories~\cite{FloryChemBook}, the results are obtained through uncontrolled approximations and rely on the cancellation of large errors 
~\cite{DeGennesBook,DesCloizeauxBook}.
While the quality of the prediction is often surprisingly good (e.g. Flory theory~\cite{GutinGrosberg93} yields $\nu = 7/13 \approx 0.54$ for self avoiding trees compared to the exact result $\nu = 1/2$ in $d=3$~\cite{ParisiSourlasPRL1981}),
it is thus crucial to complement a Flory-type analysis with results from more rigorous approaches.

The present article is the first in a series of future works~\cite{Rosa2016b,Everaers2016a,Everaers2016b,Rosa2016c},
where we use a combination of computer simulations, Flory theory and scaling arguments to investigate the connectivity and conformational statistics of randomly branched polymers with excluded volume interactions
in various ensembles.
Here we explore and compare in great detail the physical properties of three types of three-dimensional systems:
(i)
self-avoiding trees with annealed connectivity,
(ii)
self-avoiding trees with {\it quenched ideal} connectivity,
and
(iii) non-interacting trees, for which a number of analytical results exists~\cite{ZimmStockmayer49,DeGennes1968,DaoudJoanny1981} and which serve as a useful reference.
The paper is self-consistent:
following~\cite{CuiChenPRE1996,MadrasJPhysA1992,GrassbergerJPhysA2005}
we lay the groundwork, define the systems, and describe the methods for simulating them as well as for analyzing their connectivity.
While the exponents $\epsilon$, $\rho$ and $\nu$ were measured before~\cite{CuiChenPRE1996,MadrasJPhysA1992,GrassbergerJPhysA2005}, we believe to be the first to have estimated $\nu_{path}$ and $\theta$ as defined in Eqs.~(\ref{eq:nu_path}) and (\ref{eq:theta_path}).

Of our forthcoming articles:
Ref.~\cite{Rosa2016b} generalizes the tools developed here to the novel class of melt of trees in $2d$ and $3d$.
Refs.~\cite{Everaers2016a,Everaers2016b} review and generalize the Gutin {\it et al.}~\cite{GutinGrosberg93} Flory theory in the light of our present results,
those obtained in Ref.~\cite{MadrasJPhysA1992} and the above mentioned results for melts of trees.
Finally, Ref.~\cite{Rosa2016c} is devoted to an exhaustive analysis of computationally accessible distribution functions characterizing tree configurations and conformations,
which carry a wealth of information beyond what can be described by Flory theory.
To our knowledge, such an analysis has not been carried out before, additionally motivating our interest in simulating the reference systems (i-iii).

The present paper is organised as follows:
In Section~\ref{sec:theory},
we define the observables and critical exponents studied in the paper
and summarise the predictions from Flory theory. 
In Section~\ref{sec:modmethods},
we present the methods we employ to simulate the different ensembles of lattice trees, to analyse tree connectivities, and to extract the scaling behaviour of the various observables.
We present our results in Sec.~\ref{sec:results} and discuss them in Sec.~\ref{sec:Discussion}.
Finally, we sketch our conclusions in Sec.~\ref{sec:concls}.

\section{Background, model and definitions}\label{sec:theory}

We are interested in randomly branched polymers with repulsive, short-range interactions between monomers. Our choice of units and notation is explained in Sec.~\ref{sec:units}, the employed lattice model and the various ensembles are defined in Secs.~\ref{sec:LatticeModel} and \ref{sec:ensembles}, respectively.
In Section~\ref{sec:observables} we define a range of observables, which can be used to quantify the connectivity and the spatial configuration of trees. 
Known exact results for ideal trees without volume interactions are briefly summarized in section~\ref{sec:idealTrees}, while Section~\ref{sec:Flory} reviews predictions of Flory theory for interacting trees.
All our numerical results are obtained for trees embedded in $d=3$ dimensions, even though many theoretical expressions are conveniently expressed for general $d$.

\subsection{Units and notation}\label{sec:units}
We measure energy in units of $k_BT$, length in units of the Kuhn length, $l_K$, and mass in units of the number of Kuhn segments.
We use the letters $N$ and $n$ to denote the mass of a tree or a branch, respectively.
With $N$ Kuhn segments connecting the nodes of a tree, there are $N+1$ nodes in a tree.
The symbols $L$ and $l$ are reserved for contour lengths of {\em linear} paths on the tree, while $\delta L$ and $\delta l$ denote contour distances from a fixed point, typically the tree center.
Spatial distances are denoted by the letters $R$ and $r$.
Examples are the tree gyration radius, $R_g$, spatial distances between nodes, $\vec r_{ij}$, and the spatial distances, $\vec{\delta r}_i$, of a node from the tree center of mass.

\subsection{Interacting lattice trees}\label{sec:LatticeModel}
We mainly study lattice trees on the $3d$-cubic lattice. The functionality of the nodes is restricted to the values $f=1$ (a leaf or branch tip), $f=2$ (linear chain section), and $f=3$ (branch point). Connected nodes occupy adjacent lattice sites. Since our models do not include a bending energy, the lattice constant equals the Kuhn length, $l_K$, of linear paths across ideal trees. 

A tree conformation, ${\cal T}\equiv ({\mathcal G},\Gamma)$, can be described by the set of node positions, $\Gamma=\{\vec r_1, \dots, \vec r_{N+1}\}$, in the embedding space and a suitable representation of its connectivity graph, ${\mathcal G}$. We employ a data structure in the form of a linked list, which retains for each node, $i$, its position, $\vec r_i$, functionality, $f_i$, and the indices $\{j_1(i),\ldots,j_{f_i}(i)\}$ of the nodes to which it is connected. 

For ideal trees, nodes do not interact and their asymptotic branching probability, $\lambda$, is controlled via a chemical potential for branch points,
\begin{equation}\label{eq:HIdeal}
{\mathcal H}_{id}({\cal T}) =  \mu_{br} n_3({\mathcal G})
\end{equation}
where $n_3({\mathcal G})$ is the total number of  3-functional nodes in the tree.
All our results are obtained for a value of $\mu_{br} = -2.0 \, k_B T$~\cite{RosaEveraersPRL2014}.
Interactions between nodes are accounted for via
\begin{equation}\label{eq:HVolumeInteractions}
{\mathcal H}_{int}({\cal T}) =  v_K \sum_{j \in lattice} \kappa_j^2
\end{equation}
where $\kappa_j$ is the total number of Kuhn segments inside the elementary cell centered at the lattice site $j$.

\subsection{Ensembles}\label{sec:ensembles}

The statistical (Boltzmann) weight
\begin{equation}\label{eq:Boltzmann tree}
w_{\mathcal H}({\mathcal G},\Gamma)=e^{-\beta {\mathcal H}({\mathcal G},\Gamma)}
\end{equation}
of a tree conformation depends on both, the tree's connectivity, ${\mathcal G}$, and its spatial conformation, $\Gamma$. In the following, we employ a notation where probabilities, partition functions and expectation values carry a subscript denoting the Hamiltonian governing the annealed degrees of freedom and, if necessary, a superscript, characterizing the (distribution) of quenched connectivities.

\subsubsection{Annealed connectivity}

For trees with annealed connectivity, tree conformations are observed with a probability
\begin{eqnarray}
p_{\mathcal H}({\mathcal G},\Gamma) &=& \frac{w_{\mathcal H}({\mathcal G},\Gamma)}{\mathcal Z_{\mathcal H}}
\label{eq:p Boltzmann tree}\\
\mathcal Z_{\mathcal H}  &=& \sum_{\mathcal G} \int d\Gamma \,  w_{\mathcal H}({\mathcal G},\Gamma)
\label{eq:Z tree}
\end{eqnarray}
and ensemble averages of observables are given by
\begin{equation}\label{eq:A annealed}
\langle A \rangle_{\mathcal H} = \sum_{\mathcal G} \int d\Gamma \, p_{\mathcal H}({\mathcal G},\Gamma)\, A({\mathcal G},\Gamma)\ .
\end{equation}

The statistical weight of a particular tree connectivity in the annealed ensemble is given by integrating over the spatial degrees of freedom:
\begin{eqnarray}
{\mathcal Z_{\mathcal H}^{\mathcal G}}  &=&\int d\Gamma \,  w_{\mathcal H}({\mathcal G},\Gamma)
\label{eq:Z G}\\
\mathcal Z_{\mathcal H}  &=& \sum_{\mathcal G} {\mathcal Z}_{\mathcal H}^{\mathcal G}
\label{eq:Z tree bis}\\
p_{\mathcal H}({\mathcal G}) &=& \frac{{\mathcal Z}_{\mathcal H}^{\mathcal G}}{\mathcal Z_{\mathcal H}}
\label{eq:p G}
\end{eqnarray}

\subsubsection{Quenched connectivity}

As a first step, consider an ensemble of trees with annealed spatial degrees of freedom, $\Gamma$, and a given, unique connectivity, ${\mathcal G}$. A conformation is observed with a probability
\begin{eqnarray}
p_{\mathcal H}^{\mathcal G}(\Gamma) &=& \frac{w_{\mathcal H}({\mathcal G},\Gamma)}{\mathcal Z_{\mathcal H}^{\mathcal G}}
\label{eq:p Boltzmann tree bis}
\end{eqnarray}
and ensemble averages for observables are calculated from
\begin{equation}\label{eq:A G}
\langle A \rangle_{\mathcal H}^{\mathcal G} = \int d\Gamma \, p_{\mathcal H}^{\mathcal G}(\Gamma)\,A({\mathcal G},\Gamma) .
\end{equation}

In general, ensembles of trees with quenched connectivity are characterised by a given probability, $p({\mathcal G})$, for observing particular connectivities, ${\mathcal G}$. Expectation values are calculated by averaging expectation values for particular connectivities, $\langle A \rangle_{\mathcal H}^{\mathcal G}$, over the distribution $p({\mathcal G})$:
\begin{eqnarray}\label{eq:A quenched}
\langle A \rangle_{\mathcal H}^{p({\mathcal G})} 
&=& \sum_{\mathcal G} p({\mathcal G})\, \langle A \rangle_{\mathcal H}^{\mathcal G}
\end{eqnarray}

\subsubsection{Randomly quenched connectivity}

The connectivity distribution, $p({\mathcal G})$, of branched polymers is determined by the process, by which they are synthesised. Trees with ``randomly quenched connectivity'' are generated, if the connectivity of annealed trees is quenched after they have been equilibrated under the influence of some Hamiltonian, ${\mathcal H}_0$. The probability to observe a particular connectivity in the randomly quenched ensemble is then given by Eq.~(\ref{eq:p G}):
\begin{eqnarray}
p_{\mathcal H_0}({\mathcal G}) &=& \frac{{\mathcal Z}_{\mathcal H_0}^{\mathcal G}}{\mathcal Z_{\mathcal H_0}}
\end{eqnarray}
Expectation values for this ensemble have to be calculated from Eq.~(\ref{eq:A quenched}) using the Hamiltonian, ${\mathcal H}$, describing the environmental conditions at the time of observation.
\begin{eqnarray}\label{eq:A randomly quenched}
\langle A \rangle_{\mathcal H}^{\mathcal H_0} 
&=& \sum_{\mathcal G} p_{\mathcal H_0}({\mathcal G})\, \langle A \rangle_{\mathcal H}^{\mathcal G}
\end{eqnarray}
In this paper, we investigate self-avoiding trees with randomly quenched {\it ideal} connectivity.
They are generated using ${\mathcal H_0} = {\mathcal H}_{id}$ (Eq.~(\ref{eq:HIdeal})) and subsequently studied in good solvent conditions using ${\mathcal H} = {\mathcal H}_{id} + {\mathcal H}_{int}$ (Eq.~(\ref{eq:HVolumeInteractions})).

Note that Eq.~(\ref{eq:A randomly quenched}) can be simplified, if the environmental conditions at the time of observation are equal to those at the time of the connectivity quench.
This holds immediately after the quench, but might also apply at a later time, if the trees are (re)equilibrated under the original conditions.
In this case, expectation values are equal to those for trees with annealed connectivity:
\begin{eqnarray}
\langle A \rangle_{\mathcal H_0}^{\mathcal H_0}
&=& \sum_{\mathcal G}  \frac{{\mathcal Z}_{\mathcal H_0}^{\mathcal G}}{\mathcal Z_{\mathcal H_0}}
         \int d\Gamma \, \frac{w_{\mathcal H_0}({\mathcal G},\Gamma)}{\mathcal Z_{\mathcal H_0}^{\mathcal G}} 
         \,A({\mathcal G},\Gamma)
\nonumber\\
&=& \sum_{\mathcal G} 
         \int d\Gamma \, \frac{w_{\mathcal H_0}(\Gamma)}{\mathcal Z_{\mathcal H_0}}A({\mathcal G},\Gamma)
\nonumber\\
&=&\langle A \rangle_{\mathcal H_0}
\end{eqnarray}

\subsection{Tree observables}\label{sec:observables}

\subsubsection{Tree connectivity}\label{sec:connectivity_observables}

A tree is a branched structure free of loops. Its connectivity can be characterized in a number of ways.
Locally, nodes connecting Kuhn segments differ according to their functionality, $f$. Branch points have a functionality $f\ge3$, for branch tips $f\equiv1$. The numbers of branch and end points are related via
\begin{equation}
n_1 = 2 + \sum_{f=3}^\infty (f-2)n_f \, ,
\end{equation}
where $n_f$ is the total number of nodes in the tree with functionality $f$.
In particular, we have for trees with a maximal connectivity of $f=3$:
\begin{eqnarray}
n_1 &=& 2+n_3
\label{eq:n1vsn3}\\
n_2 &=& N-1-2n_3
\end{eqnarray}
The large scale structure of a tree can be analyzed in terms of the ensemble of sub-trees generated by cutting bonds. 
Removal of a bond splits a tree of weight $N$ into two smaller trees of weight
$N_< = \min(n,N-1-n)$ and $N_>=\max(n,N-1-n)$. 
We denote the corresponding probabilities for the sub-tree weight as $p_N(n)$. In particular,
\begin{equation}\label{eq:EndPoints}
\langle n_1 \rangle = 2N  \, p_N(n=0)\, .
\end{equation}
Denoting the smaller of the two remaining trees as a ``branch'', one can define a mean branch weight:
\begin{equation}\label{eq:TreeBranchSize}
\langle N_{br}(N) \rangle \equiv 2 \sum_{n=0}^{(N-1)/2} \, n  \, p_N(n) \, ,
\end{equation}
which grows as a characteristic power of the tree weight, $\langle N_{br}(N) \rangle \sim N^{\epsilon}$ (Eq.~(\ref{eq:epsilon})).

Alternatively, the tree connectivity can be analyzed in terms of the statistics of  minimal distances, $l_{ij}$, of two nodes $i,j$ along linear paths on the tree.
Introducing the probability to find two nodes at a particular contour distance, $p_N(l)$, with
\begin{equation}
p_N(l=1) = \frac{N}{N (N+1)/2}=\frac2{N+1}
\end{equation}
the corresponding average distance is given by,
\begin{equation}
\langle L(N)\rangle \equiv \sum_{l=0}^{N} l \, p_N(l) \, .
\end{equation}
Furthermore, defining the node in the middle of the longest path as the central node of a tree
we also measure
(1)
the average distance, $\langle \delta L_{center}(N) \rangle$ of nodes from the central node
\begin{equation}
\langle \delta L_{center}(N)\rangle \equiv \sum_{\delta l_{center}=0}^{N} \delta l_{center} \, p_N(\delta l_{center}) \, 
\end{equation}
where $p_N(\delta l_{center})$ is the corresponding probability distribution,
and
(2)
the average length $\langle \delta L_{center}^{max}(N) \rangle$ of the longest distance from the central node.
For ensemble averages one expects (Eq.~(\ref{eq:rho})):
\begin{equation*} 
\langle L(N) \rangle \sim \langle \delta L_{center} \rangle \sim  \langle \delta L_{center}^{max}(N) \rangle \sim N^\rho
\end{equation*}
with $\rho=\epsilon$~\cite{MadrasJPhysA1992}.

Similarly, we characterize the statistics of branches by measuring
the average branch weight, $\langle N_{br}(\delta l_{root}^{max}) \rangle$,
as a function of the {\it longest} contour distance of nodes from the branch root, $\delta l_{root}^{max}(N_{br})$.
Finally, we consider the average weight of the ``core'' of tree, $\langle N_{center}(\delta l_{center}) \rangle$,
made of segments whose distance from the central node does not exceed $\delta l_{center}$.

\subsubsection{Spatial structure}\label{sec:structure_observables}

The overall spatial extension of the tree is best described through the gyration radius and the distribution, $p_N(\vec r)$, of the distances $\vec r_{ij}\equiv \vec r_i - \vec r_j$ between nodes:
\begin{eqnarray}
\label{eq:RgDef1}
R_g^2  & \equiv & \frac1{2(N+1)^2} \sum_{i,j} \left | \vec r_{ij}  \right |^2\nonumber\\
\langle R_g^2 \rangle & = & \frac12 \int_{0}^\infty 4\pi r^2\, r^2 p_N(\vec r)\ d r
\end{eqnarray}
Alternatively, one can consider the distribution $p_N(\vec{\delta r})$ of the node positions, 
$\vec{\delta r}_{i} \equiv \vec r_i - \vec r_{CM}$, relative to the tree center of mass, $\vec r_{CM} \equiv \frac1{N+1} \sum_i \vec r_i $:
\begin{eqnarray}
\label{eq:RgDef2}
 R_g^2 & = & \frac1{(N+1)} \sum_{i} \left |  \vec{\delta r}_{i}  \right |^2\nonumber\\
\langle R_g^2 \rangle & = & \int_{0}^\infty 4\pi \delta r^2\, \delta r^2 p_N(\vec{\delta r})\ d\, \delta r
\end{eqnarray}
For ensemble averages one expects $\langle R_g^2 \rangle  \sim N^{2\nu}$ (Eq.~(\ref{eq:nu})).

The spatial conformations of linear paths on the tree can be characterized using standard observables for linear polymers.
The most general information is contained in the end-to-end distance distribution,
$p_{N}( \vec r \, | l )$ of paths of length $l$ for trees of total mass $N$.
In particular, one can extract the mean-square end-to-end distance 
\begin{eqnarray}
\label{eq:R2_path}
\langle R^2(l,N) \rangle \equiv \int_{0}^\infty 4\pi r^2\, r^2 p_{N}(\vec r\,|l)\ d r
\end{eqnarray}
and, for a given contact distance $r_c$, the end-to-end closure probability
\begin{eqnarray}
\label{eq:pc_path}
\langle p_c(l,N) \rangle \equiv \int_{0}^{r_c} 4\pi r^2\, p_{N}(\vec r\,|l)\ d r
\end{eqnarray}
of such paths, which are expected to scale as (Eqs.~(\ref{eq:nu_path}) and~(\ref{eq:theta_path})):
\begin{eqnarray*}
\langle R^2(l) \rangle  & \sim & l^{2\nu_{path}}\\
\langle p_c(l) \rangle & \sim & l^{-\nu_{path} (d+\theta_{path})}
\end{eqnarray*}
and to be asymptotically independent of tree weight. 
By construction
\begin{eqnarray}
p_N(r) &=& \int_0^{\infty} p_N(r|l) \, p_N(l) \, d l
\label{eq:p_convolution}\\
\langle R_g^2(N) \rangle &=& \frac12 \sum_{l=1}^{N} \langle R^2(l,N) \rangle\ p_N(l)\ \ .
\label{eq:Rg_R2path}
\end{eqnarray}
so that 
\begin{equation}\label{eq:nu_nupath}
\nu = \nu_{path} \, \rho\ \, .
\end{equation}

For non-interacting lattice trees with Gaussian / random walk path statistics,
\begin{eqnarray}
\label{eq:R2_path_Gauss}
\langle R^2(l) \rangle & = & l_K l\\
\label{eq:pc_path_Gauss}
p_c(l)  & \sim & l^{-d/2}
\end{eqnarray}
so that
\begin{equation}
\label{eq:Rg_R2path_ideal}
\langle R_g^2(N) \rangle = \frac12 \sum_{l=1}^{N} l_K \, l \, p_N(l) = \frac12 l_K \langle L(N)\rangle\ \, .
\end{equation}
At the same time, the Kramers theorem~\cite{RubinsteinColby} links the gyration radius to the branch weight statistics:
\begin{equation}\label{eq:TreeRg2}
\langle R_g^2(N) \rangle =
\frac{l_K^2 N}{(N+1)^2} \sum_{n=0}^{N-1} \, (n+1) \, (N-n) \,p_N(n) \, .
\end{equation}

\subsection{Ideal lattice trees}\label{sec:idealTrees}

Daoud and Joanny~\cite{DaoudJoanny1981} calculated the partition function ${\mathcal Z}_N$ of ideal lattice trees of $N$ Kuhn segments in the continuum approximation:
\begin{equation}\label{eq:DJpartfunct}
{\mathcal Z}_N = \frac{I_1 \left( 2 \, \lambda \, N \right) }{\lambda \, N} \simeq
\left\{
\begin{array}{cc}
\frac{e^{2 \lambda N}}{2 \pi^{1/2} (\lambda N)^{3/2}}, \, & \lambda N \gg 1 \\
\\
1 + \frac{(\lambda N)^2}{2} , \, & \lambda N \ll 1
\end{array}
\right.
\end{equation}
where $I_1(x)$ is the first modified Bessel function of the first kind, $\lambda$ the branching probability per node and ${\mathcal Z}_0=1$.
By employing the Kramers theorem,
\begin{equation}\label{eq:Kramers_SplitProb}
p_N(n) = \frac{ {\mathcal Z}_n \, {\mathcal Z}_{N-1-n} } { \sum_{n=0}^{N-1} \, {\mathcal Z}_n \, {\mathcal Z}_{N-1-n} } \, ,
\end{equation}
From this expression, it is possible to derive the following asymptotic relations for the quantities introduced above. 
For the branch weight distribution, 
\begin{equation}\label{eq:DaoudJoanny_SplitProb}
p_N(n) \simeq \left\{
\begin{array}{cc}
\frac{\lambda \, \left( \lambda N \right)^{3/2}}{4 \, \pi^{1/2} \left( \lambda n \right)^{3/2} \left( \lambda (N-n) \right)^{3/2}}  , \, & \lambda n \gg 1\\
\\
\frac{\lambda}{4 \, \pi^{1/2} \left( \lambda n \right)^{3/2}} , \, & \lambda N \gg \lambda n \gg 1\\
\\
\frac{1}{N} , \, & \lambda N \ll 1
\end{array}
\right. \, .
\end{equation}
For the gyration radius
\begin{equation}\label{eq:DaoudJoanny}
\langle R_g^2(N) \rangle \simeq l_K^2 \left\{
\begin{array}{cc}
\frac{\pi^{1/2}}{4 \, \lambda} \left( \lambda N \right)^{1/2} , \, & \lambda N \gg 1\\
\\
\frac{N(N+2)}{6(N+1)} , \, & \lambda N \ll 1
\end{array}
\right. \, ,
\end{equation}
in agreement with the original results of Zimm and Stockmeyer~\cite{ZimmStockmayer49}.
The average branch weight (see definition, Eq.~(\ref{eq:TreeBranchSize})) is given by
\begin{equation}\label{eq:DaoudJoanny_BranchSize}
\langle N_{br}(N) \rangle \simeq \left\{
\begin{array}{cc}
\frac{1}{(2 \pi)^{1/2} \lambda} \left( \lambda N \right)^{1/2} , \, & \lambda N \gg 1\\
\\
\frac{N^2-1}{4 N}, \, & \lambda N \ll 1
\end{array}
\right. \, 
\end{equation}
showing that $\epsilon=1/2$ for ideal trees, and the average fraction of $f=3$ nodes is given by
\begin{equation}\label{eq:brProb}
\frac{\langle n_3 \rangle}{\lambda N} \simeq
\left\{
\begin{array}{ll}
1 , \, & \lambda N \gg 1 \\
\\
\lambda N , & \lambda N \ll 1
\end{array}
\right. \, .
\end{equation}

\subsection{Flory theory}\label{sec:Flory}

Flory theories~\cite{FloryChemBook} are formulated as a balance of an entropic elastic term and an interaction energy, 
\begin{equation}\label{eq:fFlory}
{\mathcal F} = {\mathcal F_{el}(N,R)}+{\mathcal F_{inter}(N,R)} \, .
\end{equation}
$\frac{\mathcal F_{inter}(N,R)}{k_BT}=v_2 \frac{N^2}{R^d}$ represents the standard two-body repulsion between segments, which dominates in good solvent. 
Gutin {\it et al.}~\cite{GutinGrosberg93} proposed the following elastic free energy for annealed trees:
\begin{equation}\label{eq:fGutin}
\frac{{\mathcal F}_{el}}{k_B T} \sim \frac{R^2}{l_K L} +  \frac{L^2}{N \, l_K^2}  \ .
\end{equation}
The expression reduces to the entropic elasticity of a linear chain~\cite{FloryChemBook}, $\frac{{\mathcal F}_{el}(N,R)}{k_BT}=\frac{R^2}{l_K^2 N}$, 
for unbranched trees with quenched $L=l_K N$.
The first term of Eq.~(\ref{eq:fGutin}) is the usual elastic energy contribution for stretching a polymer of linear contour length $L$ at its ends.
The second term is less obvious: it is calculated from the partition function of an ideal branched polymer of $N$ bonds with $L$ bonds between two arbitrary fixed ends~\cite{GrosbergNechaev2015}.

For trees with quenched connectivity, Eq.~(\ref{eq:fGutin}) has to be evaluated for the given path length $L$ and then minimised with respect to $R$.
In particular, branched polymers with a given quenched connectivity, $L \sim N^\rho$, and mass, $N$, are expected to swell to a size of
\begin{eqnarray}
\label{eq:nu_of_rho}
\nu &=&\frac{2+\rho}{d+2}\\
\label{eq:nupath_of_rho}
\nu_{path}&=& \frac{2+\rho}{\rho(d+2)}
\end{eqnarray}
Eqs.~(\ref{eq:nu_of_rho}) and (\ref{eq:nupath_of_rho}) reduce to the classical Flory result, $\nu=\nu_{path}=3/(d+2)$, for linear chains with $\rho=1$ in $1\le d\le 4$ dimensions.
The result is exact, except in $d=3$, where $\nu=0.588$~\cite{LeGuillouZinnJustin}.
For trees with {\it quenched ideal} connectivity, $\rho=1/2$, one recovers in $d\le8$ the Isaacson and Lubensky~\cite{IsaacsonLubensky} prediction, 
\begin{eqnarray}
\label{eq:nu_of_rho IL}
\nu &=&\frac{5}{2(d+2)}\\
\label{eq:nupath_of_rho IL}
\nu_{path}&=& \frac{5}{d+2}
\end{eqnarray}

For swollen randomly branching polymers with annealed connectivity, Eq.~(\ref{eq:fGutin}) needs to be minimised with respect to both, $R$ and $L$.
In particular, Gutin {\it et al.}~\cite{GutinGrosberg93} found:
\begin{eqnarray}
\label{eq:nu_of_d}
\nu &=&\frac{7}{3d+4}\\
\label{eq:nupath_of_d}
\nu_{path}&=& \frac{7}{d+6}\\
\label{eq:rho_of_d}
\rho&=& \frac{d+6}{3d+4}
\end{eqnarray}
In $d=1$, the chain is predicted to be unbranched and fully stretched, $\nu=\rho = \nu_{path}=1$.
In $2\le d \le 4$ the predicted values are in reasonable agreement with numerical results~\cite{MadrasJPhysA1992},
even though the prediction $\nu = 7/13 \approx 0.54$ slightly exceeds the exact result $\nu = 1/2$ in $d=3$ \cite{ParisiSourlasPRL1981}.
In particular, in $d=3$ $\nu_q = 1/2 < \nu_a = 7/13$ and $\nu_{path, q} = 1 > \nu_{path, a} = 7/9$.
Consequently, annealed and quenched trees belong to different universality classes~\cite{GutinGrosberg93}.
Notice, that the Flory theory gives no prediction for the exponent $\theta_{path}$ of contact probabilities between tree nodes, Eq.~(\ref{eq:theta_path}).

In the following, we will compare these results to computer simulations of trees in the two different ensembles.
The same ideas can also be applied to a tree melt:
corresponding results will be presented and discussed in forthcoming publications~\cite{Rosa2016b,Everaers2016a,Everaers2016b,Rosa2016c}.

\section{Models and methods}\label{sec:modmethods}

In this section, we describe the algorithms used for simulating trees with annealed (Sec.~\ref{sec:AmoebaAlgo}) and quenched (Sec.~\ref{sec:MDmethods}) connectivity,
for analysing the connectivity of branched polymers (Sec.~\ref{sec:burning}) and for estimating asymptotic exponents (Sec.~\ref{sec:ExtractExps}).
Quantitative details as well as tabulated values for single-tree statistics are reported in Supplementary Data.

\begin{figure}
\begin{center}
\includegraphics[width=0.48\textwidth]{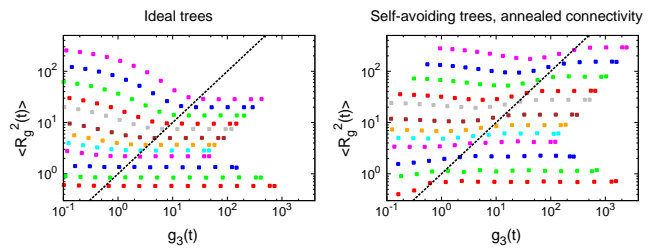}
\end{center}
\caption{
\label{fig:MC_Equilibration}
Parametric plots of the MC-time ($t$) evolution
of the ensemble-average square gyration radius, $\langle R_g^2(t) \rangle$,
{\it vs.} the mean-square displacement of the tree center of mass, $g_3(t)$.
Non-equilibrated (resp., equilibrated) values of the plots correspond
to regions above (resp., below) the black solid line $y=x$.
Different colors correspond to different tree masses ranging from $N=5$ (bottom) to $N=1800$ (top).
}
\end{figure}

\subsection{Monte Carlo simulations of lattice trees with annealed connectivity}\label{sec:AmoebaAlgo}

We use the lattice model defined in Sec.~\ref{sec:LatticeModel} to study trees with annealed connectivity, because the corresponding ensembles can be conveniently generated with the help of the ``amoeba'' Monte Carlo algorithm by Seitz and Klein~\cite{SeitzKlein1981}. 

All simulations are carried out on a $3d$-cubic lattice with periodic boundary conditions.
We adopt a large box, so that the average density of Kuhn segments per cell is around a few percent.
As in Ref.~\cite{RosaEveraersPRL2014}, we employ a large free energy penalty of $v_k = 4 \, k_B T$ for overlapping pairs of Kuhn segments, Eq.~(\ref{eq:HVolumeInteractions}).
The pair repulsion is so strong, that single trees are effectively self-avoiding. 

Amoeba trial moves simultaneously modify the tree connectivity, ${\mathcal G}$, and the tree conformation, $\Gamma$.
They are constructed by randomly cutting a leaf (or node with functionality $f=1$) from the tree and placing it on a randomly chosen site adjacent to a randomly chosen node with functionality $f<3$, to which the leave is then connected.
Trial moves, ${\cal T}_i \rightarrow {\cal T}_f$, are accepted with probability:
\begin{equation}\label{eq:accRatioIdeal}
\mbox{acc}_{i \rightarrow f} =
\min \left\{ 1, \frac{n_1(i)}{n_1(f)}
e^{-\beta \left( {\mathcal H}({\cal T}_f) - {\mathcal H}({\cal T}_i)\right)} \right\}
\end{equation}
where $n_1(i/f)$ is the total numbers of 1-functional nodes in the initial/final state
and ${\mathcal H}({\cal T}) = {\mathcal H}_{id}({\cal T}) + {\mathcal H}_{int}({\cal T})$, Eqs.~(\ref{eq:HIdeal}) and~(\ref{eq:HVolumeInteractions}).
It should be noted, that our version of the amoeba algorithm is slightly modified with respect to the original one of Ref.~\cite{SeitzKlein1981} as we impose node functionalities $f\leq 3$~\cite{RosaEveraersPRL2014}.

We have generated ideal lattice trees 
and single trees with volume interactions for tree weights of $N=3,\ldots,1800$ Kuhn segments and starting from linearly connected random walks as initial states.
As illustrated by Fig.~\ref{fig:MC_Equilibration}, the tree gyration radii equilibrate over a time scale during which the tree centers of mass diffuse over the corresponding distance. 
The total computational effort for these simulations as a function of system size is summarised in Table~S1.

\subsection{Molecular Dynamics simulations of trees with quenched connectivity}\label{sec:MDmethods}

\begin{figure}
\begin{center}
\includegraphics[width=0.48\textwidth]{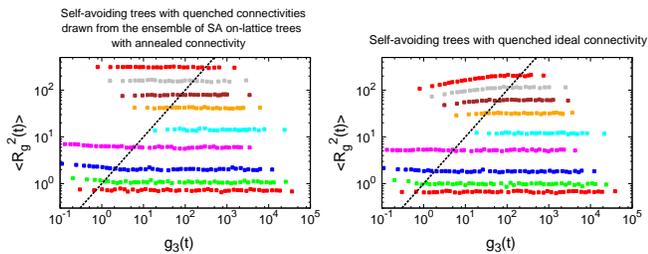}
\end{center}
\caption{
\label{fig:MD_Equilibration}
Parametric plots of the MD-time ($t$) evolution
of the ensemble-average square gyration radius, $\langle R_g^2(t) \rangle$,
{\it vs.} the mean-square displacement of the tree center of mass, $g_3(t)$.
Non-equilibrated (resp., equilibrated) values of the plots correspond
to regions above (resp., below) the black solid line $y=x$.
Different colors correspond to different tree masses ranging from $N=5$ (bottom) to $N=1800$ (top).
}
\end{figure}

In addition, we have studied randomly branched trees with quenched ideal connectivity. In this case, we have employed an equivalent off-lattice model and studied it via Molecular Dynamics simulations with the LAMMPS package~\cite{lammps}. 

As suggested by Gutin {\it et al.}~\cite{GutinGrosberg93} and explained in Sec.~\ref{sec:Flory} randomly branched trees with {\em quenched} connectivity are expected to react differently to a change of environmental conditions than randomly branching trees with {\em annealed} connectivity.
By construction, there is no difference between the two ensembles as long as the conditions equal those under which annealed degrees of freedom are quenched.
Ideally, one would investigate the two situations for the same (lattice) model.
Since amoeba moves simultaneously change node positions and tree connectivity, this would have required the implementation of a different Monte Carlo algorithm.

Instead we decided to use Molecular Dynamics simulations to explore the conformational statistics of a corresponding off-lattice, bead-spring model with the same number of degrees of freedom as the lattice model, fixed connectivity, and (approximately) identical conformational statistics for a given connectivity. The convenient one-to-one mapping facilitates the import of starting states from the lattice simulations. This allows us to study ensembles of trees whose connectivity is quenched in configurations with well-defined statistical properties.

In the bead-spring off-lattice model, bonds are modelled as harmonic springs with a rest length corresponding to the Kuhn length (or lattice constant of the lattice model).
The corresponding sum in the Hamiltonian has to run over the list of bonds specifying the tree quenched connectivity:
\begin{eqnarray}\label{eq:HarmonicBond}
{\cal H}_{bond} & = & \frac12 \sum_{i=1}^{N} \sum_{\alpha=1}^{f_i} K (r_{i,j_{\alpha}(i)}-l_K)^2
\end{eqnarray}
where $r_{kl}\equiv |\vec r_{k}-\vec r_l|$ and where we have double counted all bonds to ease the notation. For a linear chain, $f_1=1$ and $j_1(1)=2$, $f_i=2$ with $j_1(i)=i-1$ and $j_2(i)=i+1$ $\forall i \in ]1,N[$, and $f_N=1$ and $j_1(N)=N-1$ so that 
$
{\cal H}_{bond}  =  \sum_{i=1}^{N-1} K (r_{i+1,i}-l_K)^2
$.
We use a spring constant  $K / k_B T = 30/l_K^2$.

Furthermore, beads interact with a soft repulsive potential
\begin{eqnarray}\label{eq:SoftPot}
{\cal H}_{EV} & = & \sum_{i=1}^{N-1} \sum_{j=i+1}^{N} V(r_{ij})
\end{eqnarray}
where
\begin{equation}\label{eq:SoftPot1}
V(r) = \left \{
\begin{array}{cc}
A_K \left[ 1+\cos\left(\frac{\pi \, r}{1.12 \, l_K} \right)  \right] , & r \leq 1.12 \, l_K \\
\\
0 , & \mbox{otherwise}
\end{array}
\right.
\end{equation}
The range of the function $V(r)$ corresponds to the choice usually employed in the standard Kremer-Grest model~\cite{KremerGrestJCP1990} with pairwise Lennard-Jones repulsive interactions.
$A_K = 1.0 \, k_B T$ controls the strength of the excluded volume interactions similarly to $v_K$ in the lattice model Eq.~(\ref{eq:HVolumeInteractions}).
This value was tuned on the basis that the gyration radii of self-avoiding lattice trees must not change after switching to the off-lattice model, if their quenched connectivities are drawn from the ensemble of self-avoiding on-lattice trees with annealed connectivity. 
As illustrated by Fig.~\ref{fig:MD_Equilibration}, gyration radii remain virtually unchanged over the course of simulations, which are long enough that trees can move over distances much larger their own sizes.
On the other hand, MD simulations initialized with configurations of ideal lattice trees
show that ensemble-average gyration radii increase with time,
saturating to asymptotic values which are typically {\it smaller} than the ones obtained in the previous case.

Equilibrated values of mean-square gyration radii of trees in the two ensembles including details on the computational cost of MD simulations
are summarised in Table~S2.

\subsection{Connectivity  analysis via ``burning''}\label{sec:burning}

\begin{figure}
\begin{center}
\includegraphics[width=0.45\textwidth]{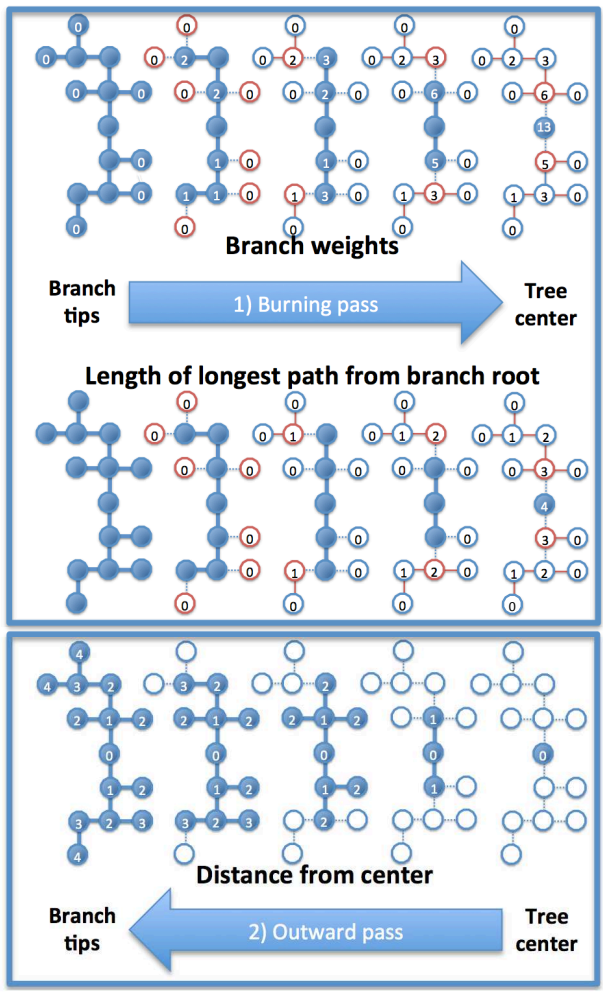}
\end{center}
\caption{
\label{fig:Burning}
Illustration of the burning algorithm applied to a tree of $N=13$ Kuhn segments.
The top and center panel illustrate the information on the branches obtained in the initial inward pass, which starts from the branch tips and locates the tree center.
The bottom panel illustrates the outward pass, which allows to determine the distance of nodes from the tree center.
}
\end{figure}

We have analyzed tree connectivities using a variant of the ``burning'' algorithm for percolation clusters~\cite{StanleyJPhysA1984,StaufferAharonyBook}.
The algorithm is very simple, and consists of two parts, see Fig.~\ref{fig:Burning}.
In the initial inward (or burning) pass branch tips are iteratively ``burned'' until the tree center is reached. In the subsequent outward pass one advances from the center towards the periphery. The inward pass provides information about the mass and shape of branches. The outward pass allows to reconstruct the distance of nodes from the tree center.

Cutting a segment between two nodes splits a tree of weight $N$ into two smaller trees of weight $n$ and $N-1-n$ (see section~\ref{sec:connectivity_observables}).
The burning algorithm iteratively removes the segments from the outside to the tree center and associates the information about branches with the root node via which they used to be connected to the tree.
The procedure is initialized by setting the branch masses, $n_i$, of all nodes to zero.
In the first iteration of the inward pass, one removes or burns the segments connecting the tree to the original branch tips.
The resulting branches consist of single nodes. Their mass, $n_i=0$, is properly set by the initialization.
The longest paths from the root node on these branch have a length of $\delta l_{root}^{max}=0$.
For bookkeeping purposes the sum of the masses of the removed branches and of the removed segments, $n_i+1$,
is added to the mass of the tree nodes to which the branches used to be connected.
In each subsequent iteration of the burning algorithm the same procedure is applied to the $1$-functional nodes of the remaining tree.
The bookkeeping scheme automatically records branch weights ({\it i.e.} the number of segments connected to the tree through a given node) before the root node of the branch is cut off. For the branches cut in the $j$-th iteration, the longest path from the root node has a length of $\delta l_{root}^{max}=j-1$.
The procedure stops, when a single node or a single segment remain.
In the former case, the accumulated weight on the remaining, central node equals the tree weight.
In the latter case, we designate one of the two remaining nodes as being the central node before burning the last segment.
Note that branch weights $n_i$ recorded in the core of the tree can exceed half the tree weight.
When binning the branch weight histogram we therefore use the $n_{i<}=\min(n_i,N-1-n_i)$. 

By construction, the central node is located in the middle of the longest path on the tree. The outward pass of the algorithm is trivial. Starting from the central node, one advances at each iteration to the outward neighbors of the nodes considered in the previous iterations. In iteration $j$ their distance to the center is given by $\delta l_{center}=j$.

\subsection{Extracting exponents from data for finite-size trees}\label{sec:ExtractExps}

The discussion in Sec.~\ref{sec:theory} applies to the asymptotic behaviour of large trees.
In this limit, estimates of the various critical exponents
($\rho, \epsilon, \nu_{path}, \nu$)
can be obtained by plotting numerical results for quantities such as
the mean-square gyration radius, $\langle R_g^2(N) \rangle$, in a log-log plot, and fitting the data by linear regression to a straight line:
\begin{equation}\label{eq:FitRg22}
\log \langle R_g^2(N) \rangle = a + 2 \nu \log N \, .
\end{equation}
Similar expressions can be written for all quantities analyzed here, namely:
$\langle \delta L_{center}(N) \rangle$,
$\langle \delta L_{center}^{max}(N) \rangle$,
$\langle L(N) \rangle$, and
$\langle N_{br}(N) \rangle$
with corresponding critical exponents.
Eq.~(\ref{eq:FitRg22}) corresponds to pure power-law behavior where the sought critical exponent is inferred from the slope of the line
determined by minimizing $\chi^2 = \sum_{i=1}^{D} \left[ \frac{\log \langle R_g^2(N_i)\rangle_{observed} - \log \langle R_g^2(N_i)\rangle_{model}}{\delta \log \langle R_g^2(N_i) \rangle} \right]^2$, with $D$ the number of data points used in the fit.

For finite tree sizes corrections-to-scaling induce systematic errors in the result.
To include them into the error analysis we follow a procedure inspired by Ref.~\cite{MadrasJPhysA1992}, which combines two different analysis schemes. 
The first is based on Eq.~(\ref{eq:FitRg22}). In the presence of corrections to scaling extracted ``effective'' exponents depend on the window of tree sizes included in the fit. To obtain results as close as possible to the asymptotic limit, our fits to Eq.~(\ref{eq:FitRg22}) are exclusively based on data for the three largest available tree sizes, $N = 450, 900,1800$.  In this scheme, no attempt is made to extrapolate the results beyond the range of studied tree sizes. 

The second scheme, again applied to quantities plotted in a log-log plot, uses the functional form
\begin{equation}\label{eq:FitRg21}
\log \langle R_g^2(N) \rangle = a + b N^{-\Delta} + 2 \nu \log N \, ,
\end{equation}
for a power-law with a {\em single} correction-to-scaling term.
Since Eq.~(\ref{eq:FitRg21}) proposes to describe the deviations from the asymptotic behavior,
our fits to this expression take into account data for all tree sizes with $N \geq 10$.
{\it A priori} Eq.~(\ref{eq:FitRg21}) poses a non-linear optimization problem in a four dimensional parameter space. 
By minimizing  $\chi^2(\Delta)$ over a suitable range of values for $\Delta$,
one can reduce the problem to a combination of a generalized linear least square fit~\cite{NumericalRecipes}
for the parameters $a$, $b$ and $\nu$ and a non-linear $1d$ optimization for the parameter $\Delta$.  
To include the uncertainties in all four parameters on an equal footing,
we have instead optimized a self-consistent linearization of Eq.~(\ref{eq:FitRg21}):
\begin{eqnarray}\label{eq:FitRg21lin2}
\lefteqn{\log \langle R_g^2(N) \rangle  =} \nonumber\\
&& a + b N^{-\Delta_0 } + 2 \nu \log N   - b  (\Delta-\Delta_0) N^{-\Delta_0 } \log N \ , \nonumber\\
\end{eqnarray}
{\it i.e.} we have carried out a one-dimensional search for the value of $\Delta_0$, for which the fit yields a vanishing $N^{-\Delta_0 } \log N$ term.

Quality of the fit is estimated by the normalized ${\tilde \chi}^2 \equiv \frac{\chi^2}{D-f}$,
where $D - f$ is the difference between the number of data points, $D$, and the number of fit parameters, $f$.
Here $f=2$ for Eq.~(\ref{eq:FitRg22}) and $f=4$ for Eq.~(\ref{eq:FitRg21}). 
When ${\tilde \chi}^2 \approx 1$ the fit is deemed to be reliable~\cite{NumericalRecipes}.
The corresponding $\mathcal Q(D-f, \chi^2)$-values provide a quantitative indicator for the likelihood that $\chi^2$ should exceed the observed value, if the model were correct~\cite{NumericalRecipes}.
All fit results are reported together with the corresponding errors, ${\tilde \chi}^2$ and $\mathcal Q$ values. 
Final estimates of critical exponents are calculated as averages of all independent measurements.
Corresponding uncertainties are given in the form
$\pm$(statistical error)$\pm$(systematic error),
where the ``statistical error'' is the largest value obtained from the different fits~\cite{MadrasJPhysA1992} while the ``systematic error'' is the spread between the single estimates, respectively.
In those cases where Eq.~(\ref{eq:FitRg21lin2}) fails producing trustable results we have retained only the $2$-parameter fit, Eq.~(\ref{eq:FitRg22}),
and a separate analysis of uncertainties was required, see the caption of Table~S6 
for details.
Error bars reported in Table~\ref{tab:ExpSummary} are given by
$\sqrt{(\mbox{statistical error})^2 + (\mbox{systematic error})^2}$.

\section{Results}\label{sec:results}

In the following sections,
we provide results concerning the structure of trees (average values of the specific observables defined in Sec.~\ref{sec:observables}) for different statistical ensembles,
as well as the critical exponents characterising the tree behaviour in the large-$N$ limit.
A complete study dedicated to melt of trees and distribution functions will be presented in forthcoming publications~\cite{Rosa2016b,Rosa2016c}.

\subsection{Branching statistics for trees with annealed connectivity}\label{sec:BranchStat}

\begin{figure}
\begin{center}
\includegraphics[width=0.5\textwidth]{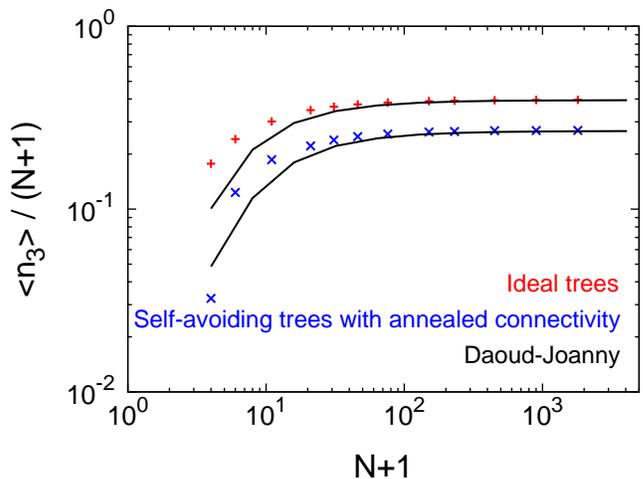}
\end{center}
\caption{
\label{fig:BranchNumber}
Branching statistics.
Average fraction of $3$-functional nodes, $\frac{\langle n_3 \rangle}{N+1}$, as a function of the total number of tree nodes, $N+1$.
Black solid lines correspond to the analytical expression for ideal trees, Eq.~(\ref{eq:brProb}),
with corresponding asymptotic branching probabilities $\lambda = 0.4$ (ideal trees) and $\lambda = 0.269$ (annealed self-avoiding trees).
}
\end{figure}

Our results for the average number of branch points, $\langle n_3(N) \rangle$, as a function of $N$ are listed in Table~S3. 
Figure~\ref{fig:BranchNumber} shows that the ratios of 3-functional nodes, $\langle n_3(N) \rangle /N$,
reach their asymptotic value already for moderate tree weights.
Our results for ideal trees perfectly agree with Eq. (\ref{eq:brProb}) for $\lambda = \lim_{N\rightarrow\infty} \langle n_3(N) \rangle /N = 0.4$, providing a first validation of our implementation of the amoeba MC method.
In contrast, we find for isolated self-interacting trees $\lim_{N\rightarrow\infty} \langle n_3(N) \rangle /N \approx 0.269$.
The value is in good agreement with the value $\approx 0.256$~\cite{MadrasJPhysA1992} for lattice trees with no restriction on node functionality.
Corresponding distributions $p(n_3)$ are well described by Gaussian statistics
with corresponding variances increasing linearly with $N$, see Fig.~S1.

\subsection{Path length statistics for trees}\label{sec:ppStat}

\begin{figure}
\begin{center}
\includegraphics[width=0.5\textwidth]{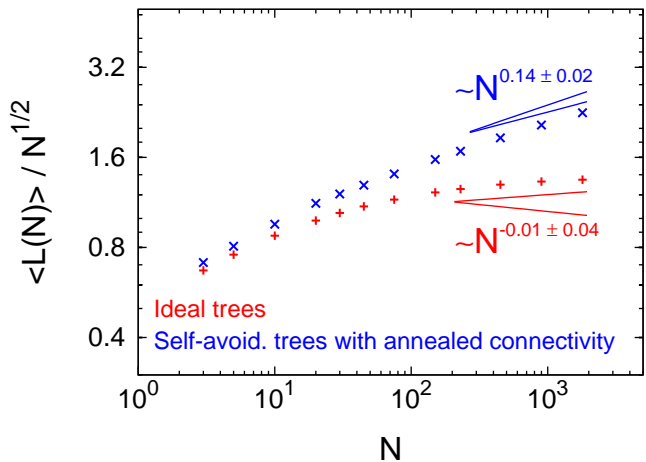}
\end{center}
\caption{
\label{fig:PathLengthStat_L}
Path length statistics.
Mean contour distance between pairs of nodes, $\langle L(N) \rangle$.
Straight lines correspond to the large-$N$ behaviour $\langle L(N) \rangle \sim N^{\rho}$ with critical exponents $\rho$
given by the best estimates summarised in Table~\ref{tab:ExpSummary}.
}
\end{figure}

Our results for 
(A)
the mean contour distance between pairs of nodes, $\langle L(N) \rangle$,
(B)
the mean contour distance of nodes from the central node, $\langle \delta L_{center}(N) \rangle$, and
(C)
the mean {\it longest} contour distance of nodes from the central node, $\langle \delta L_{center}^{max}(N) \rangle$
are summarized in Table~S3 
and plotted in Fig.~\ref{fig:PathLengthStat_L} and Fig.~S2. 
As discussed in the theory section~\ref{sec:connectivity_observables}, the three quantities are expected to scale with the total tree weight $N$ as
$\langle \delta L_{center}(N) \rangle \sim \langle \delta L_{center}^{max}(N) \rangle \sim \langle L(N) \rangle \sim N^{\rho}$.
Extracted single values for $\rho$'s including more details on their statistical significance ($\chi^2$ and $\mathcal Q$-values) are summarized in Table~S4. 
Our final best estimates for $\rho$'s (straight lines in Fig.~\ref{fig:PathLengthStat_L} and Fig.~S2, 
and Tables~\ref{tab:ExpSummary} and~S4) 
are obtained by combining the corresponding results for $\langle \delta L_{center}(N) \rangle$, $\langle \delta L_{center}^{max}(N) \rangle$ and $\langle L(N) \rangle$.

\subsection{Path lengths vs. weights of branches}\label{sec:ppStatBranches}

\begin{figure*}
\includegraphics[width=\textwidth]{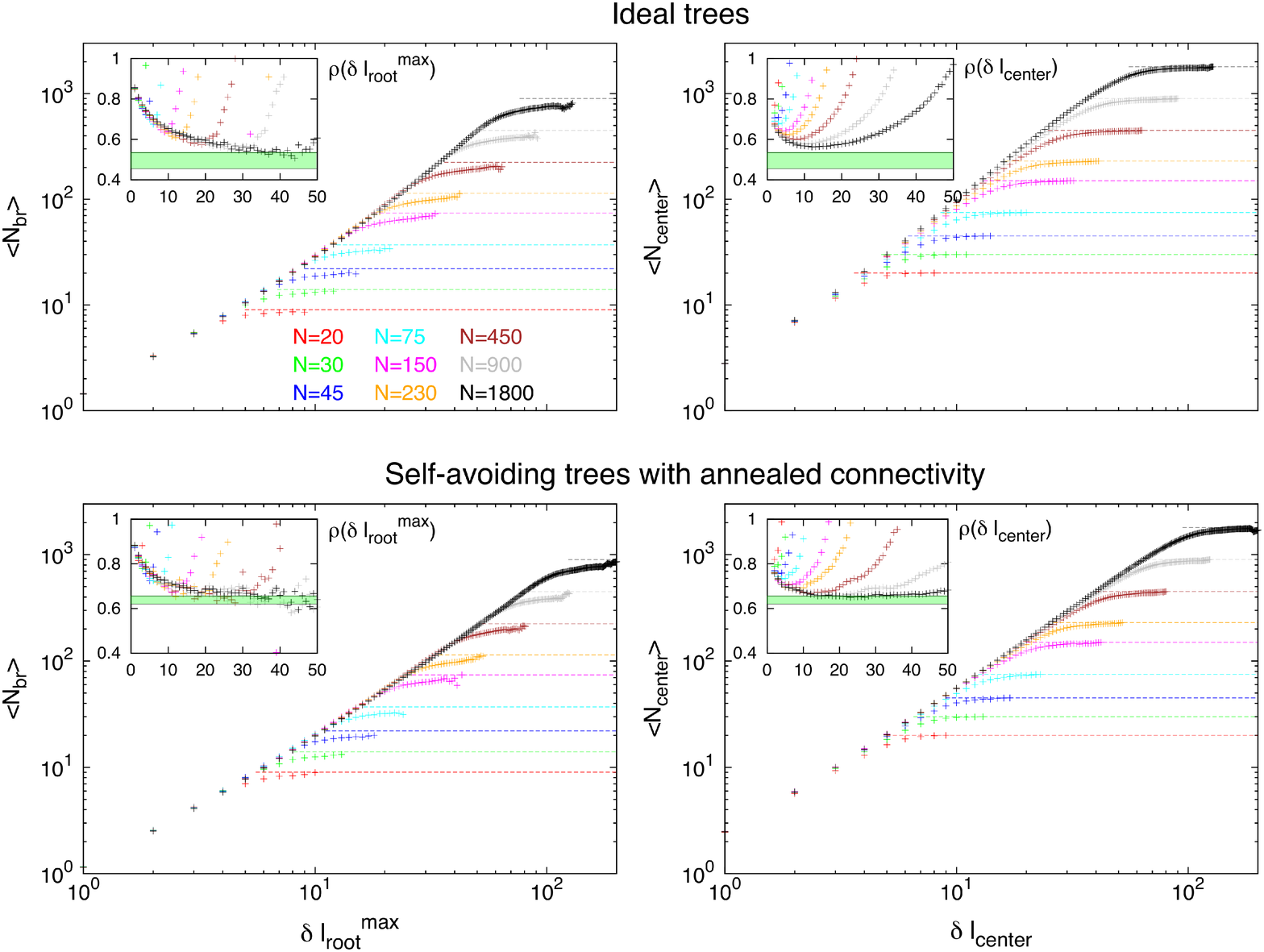}
\caption{
\label{fig:avMass_VS_lbranch}
Path lengths vs. weights of branches.
Data for trees of weight from $N=20$ to $N=1800$.
(Left)
Average branch weight, $\langle N_{br}(\delta l_{root}^{max})\rangle$, as a function of the longest contour distance to the branch root, $\delta l_{root}^{max}$.
For large $\delta l_{root}^{max}$, curves saturate to the corresponding maximal branch weight $(N-1)/2$ (dashed horizontal lines).
(Right)
Average branch weight, $\langle N_{center}(\delta l_{center})\rangle$, composed of segments whose distance from the central node does not exceed $\delta L_{center}$.
For large $\delta l_{center}$, curves saturate to the corresponding total tree weight, $N$ (dashed horizontal lines).
Insets:
Corresponding differential fractal exponent $\rho(\delta l_{root}^{max})$ and $\rho(\delta l_{center})$.
Shaded regions show the range of $\rho$ values summarized in Table~\ref{tab:ExpSummary}.
}
\end{figure*}

The relation between branch weight and path length can also be explored on the level of branches.
We have analyzed the scaling behavior of:
(1) the average branch weight, $\langle N_{br}(\delta l_{root}^{max})\rangle$, as a function of the longest contour distance to the branch root, $\delta l_{root}^{max}$, and
(2) the average branch (or tree core) weight, $\langle N_{center}(\delta l_{center})\rangle$, inside a contour distance $\delta l_{center}$ from the central node of the tree.
Corresponding results are shown in Fig.~\ref{fig:avMass_VS_lbranch}.
Both data sets show universal behavior at intermediate $\delta l_{root}^{max}$ and $\delta l_{center}$,
and saturate to the corresponding expected limiting values $=\frac{N-1}{2}$ and $=N$.
For large $\delta l_{root}^{max}$ (respectively, $\delta l_{center}$) the relation $\langle N_{br}(\delta l_{root}^{max}) \rangle \sim {\delta l_{root}^{max}}^{1/\rho}$
(resp., $\langle N_{center} (\delta l_{center}) \rangle \sim {\delta l_{center}}^{1/\rho}$) is expected to hold.
For $N_{br}$ (and with an analogous expression for $N_{center})$, we have estimated $\rho$ as
$\rho(\delta l_{root}^{max}) = \left( \frac{\log \, \langle N_{br}(\delta l_{root}^{max}+1) \rangle / \langle N_{br}(\delta l_{root}^{max}) \rangle}{\log \, (\delta l_{root}^{max}+1)/\delta l_{root}^{max}} \right)^{-1}$.
Numerical results are reported in the corresponding insets of Fig.~\ref{fig:avMass_VS_lbranch},
the large-scale behaviour agreeing well with the best estimates for $\rho$'s (shaded areas) summarized in Table~\ref{tab:ExpSummary}.

\subsection{Branch weight statistics}\label{sec:BWeightStat}

\begin{figure}
\begin{center}
\includegraphics[width=0.5\textwidth]{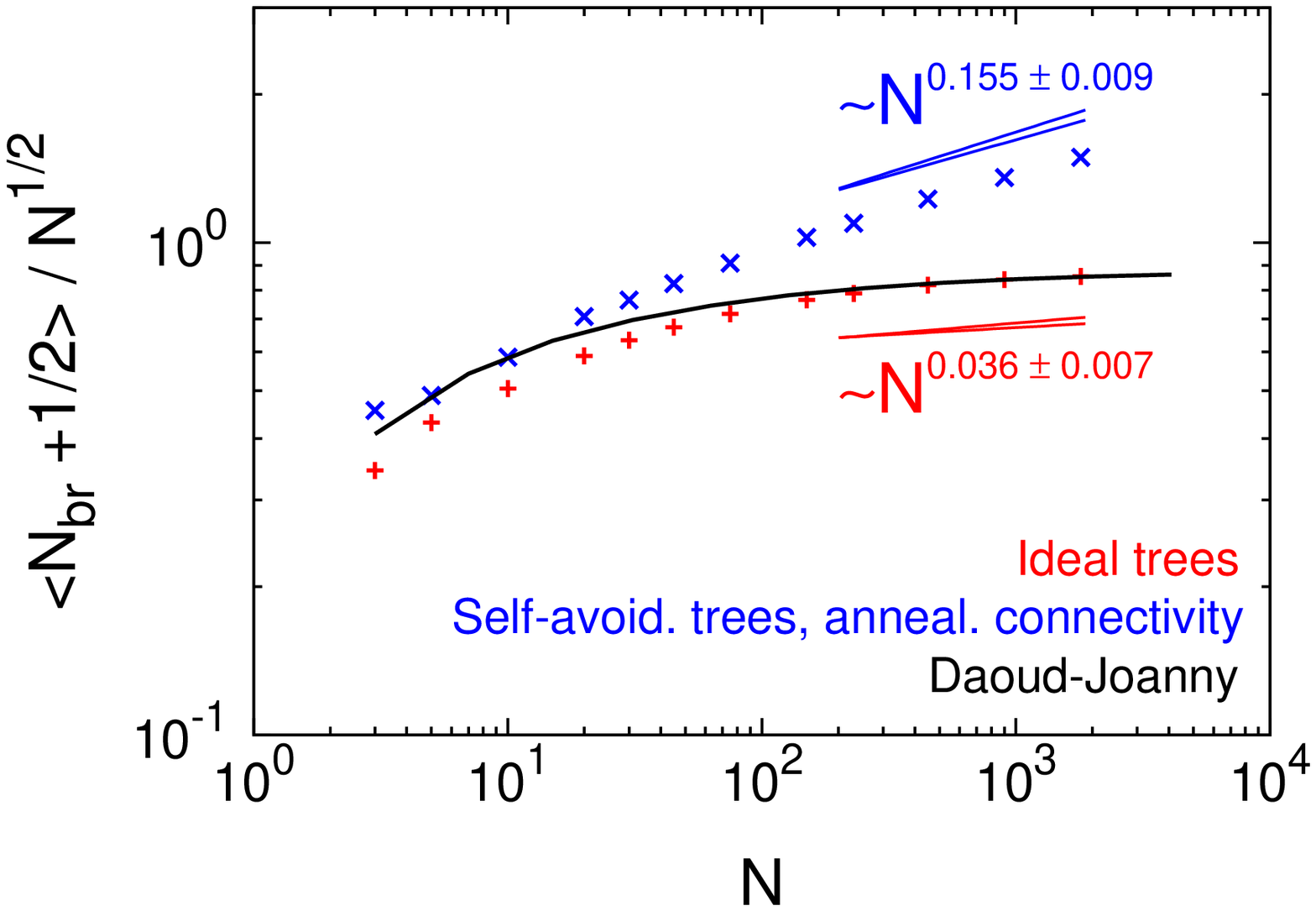}
\end{center}
\caption{
\label{fig:nBranch}
Branch weight statistics.
Average branch weight, $\langle N_{br} (N) \rangle$ as a function of the total tree mass, $N$.
Deviating from the convention used in the rest of the article, we have added half of the weight of the severed bond to the branch weights to reduce the deviations from the (continuum) theory (Eq.~(\ref{eq:DaoudJoanny_BranchSize}), black line).
Straight lines correspond to the large-$N$ behaviour $\langle N_{br} (N) \rangle \sim N^{\epsilon}$ with critical exponents $\epsilon$
given by the best estimates summarised in Table~\ref{tab:ExpSummary}.
}
\end{figure}

The scaling behavior of the average branch weight, $\langle N_{br} (N) \rangle \sim N^{\epsilon}$, defines the critical exponent $\epsilon$.
Single values of $\langle N_{br} (N) \rangle$ for each $N$ (see Table~S3) 
are plotted in Fig.~\ref{fig:nBranch},
where the straight lines have slopes corresponding to our best estimates for $\epsilon$'s (Table~\ref{tab:ExpSummary}), see also Table~S4 
for details.
We notice, in particular, that the scaling relation $\rho = \epsilon$ holds within error bars.

\subsection{Conformational statistics of linear paths}\label{sec:ConfStatPaths}

\begin{figure*}
\begin{center}
\includegraphics[width=0.7\textwidth]{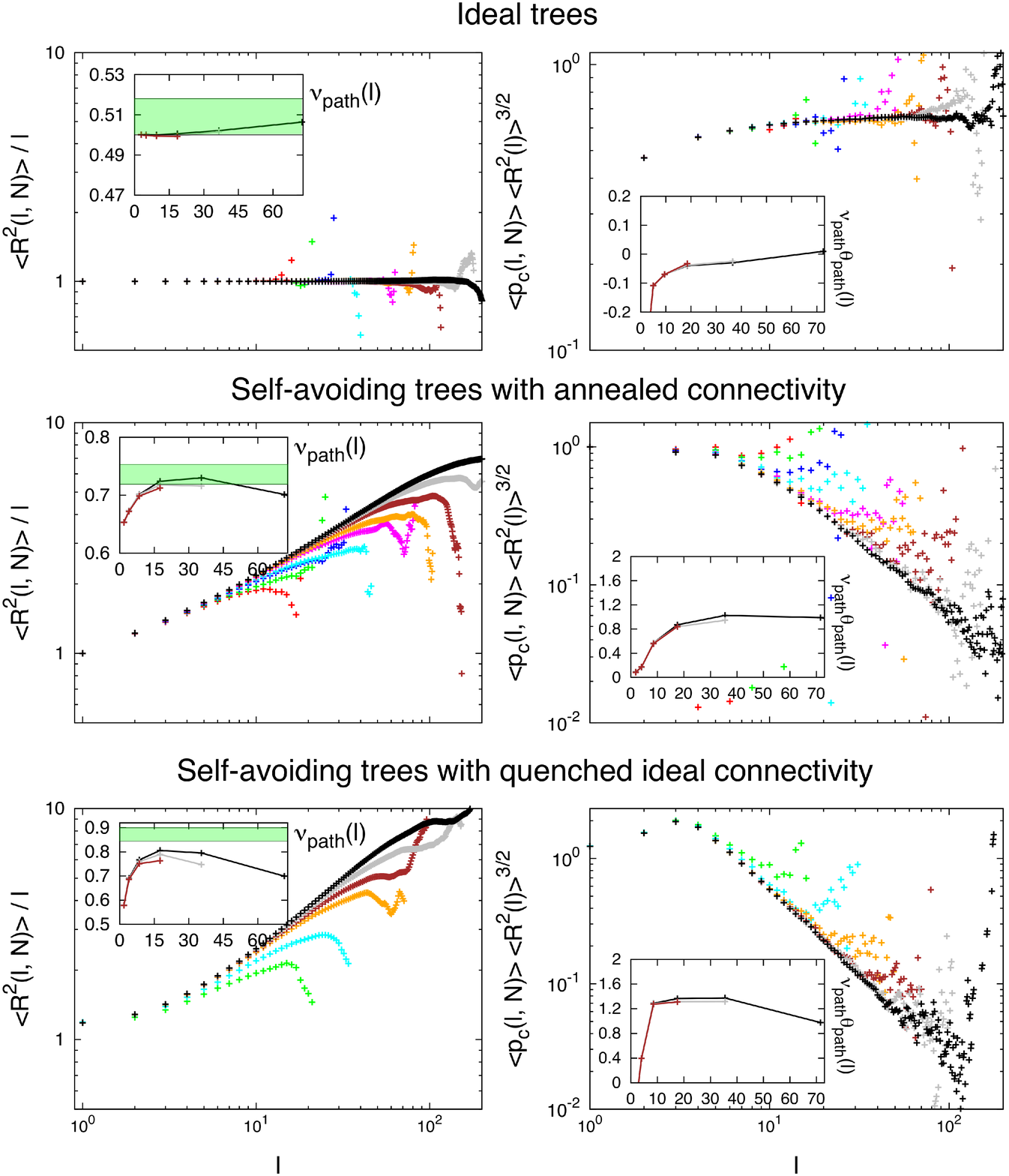}
\end{center}
\caption{
\label{fig:PathR2_ContactFreq}
Conformational statistics of linear paths.
(L.h. column)
Mean-square end-to-end distance, $\langle R^2(l, N) \rangle$, of linear paths of length $l$.
Insets:
Differential fractal exponent,
$\nu_{path}(l) = \frac{1}{2}\frac{\log \, \langle R^2(l+1, N) \rangle / \langle R^2(l, N) \rangle}{\log \, (l+1)/l}$ for chain lengths $N\geq450$.
Shaded regions show the range of $\nu_{path}$ values summarized in Table~\ref{tab:ExpSummary}.
(R.h. column)
Mean closure probabilities, $\langle p_c(l, N) \rangle$, between ends of linear paths of length $l$ normalised to the mean-field expectation value $\langle R^2(l, N) \rangle^{-3/2}$.
Insets:
Differential fractal exponent $\nu_{path} \theta_{path}(l)$, see Eq.~(\ref{eq:theta_path}), defined analogously to $\nu_{path}(l)$ for chain lengths $N\geq450$.
Plots in the insets have been obtained by averaging corresponding quantities over log-spaced intervals.
Color code is as in Fig.~\ref{fig:avMass_VS_lbranch}.
}
\end{figure*}

First, we considered the mean-square end-to-end distances, $\langle R^2(l, N) \rangle$ (Fig.~\ref{fig:PathR2_ContactFreq}, left-hand panels),
of linear paths of contour length $l$.
In order to extract the critical exponent $\nu_{path}$ which defines the scaling behavior $\langle R^2(l, N) \rangle \sim l^{2\nu_{path}}$,
we have selected paths of length equal to the average length ($l=\langle L(N) \rangle$) and to the trees maximal length ($l=L_{max}(N)$)
and calculated corresponding mean-square end-to-end distances $\langle R^2(\langle L \rangle) \rangle \sim \langle L(N) \rangle^{2\nu_{path}}$ and $\langle R^2(L_{max}) \rangle \sim \langle L_{max}(N) \rangle^{2\nu_{path}}$
(see Fig.~S3 
and corresponding tabulated values in Table~S5).
Combination of the two (Table S6) led to our best estimates for $\nu_{path}$ in the different ensembles summarized in Table 1.
Not surprisingly, these values agree well with the differential exponents
$\nu_{path}(l) = \frac{1}{2}\frac{\log \, \langle R^2(l+1, N) \rangle / \langle R^2(l, N) \rangle}{\log \, (l+1)/l}$ reported in the l.h.s insets of Fig.~\ref{fig:PathR2_ContactFreq}.

Then, we have calculated the mean closure probabilities, $\langle p_c(l, N) \rangle$ (Fig.~\ref{fig:PathR2_ContactFreq}, right-hand panels),
normalised to the corresponding ``mean-field'' expectation values $\langle R^2(l, N) \rangle^{-3/2} \sim l^{-3\nu_{path}}$.
Interestingly, $\langle p_c(l, N) \rangle$ for interacting trees markedly deviate from the mean-field prediction,
which defines a {\it novel} critical exponent $\theta_{path}$,
$\langle p_c(l, N) \rangle \langle R^2(l, N) \rangle^{3/2} \sim l^{-\nu_{path}\theta_{path}}$.
Estimated values for $\theta_{path}$'s are reported in Table~\ref{tab:ExpSummary}.

\subsection{Conformational statistics of trees}\label{sec:ConfStatTrees}

\begin{figure}
\begin{center}
\includegraphics[width=0.5\textwidth]{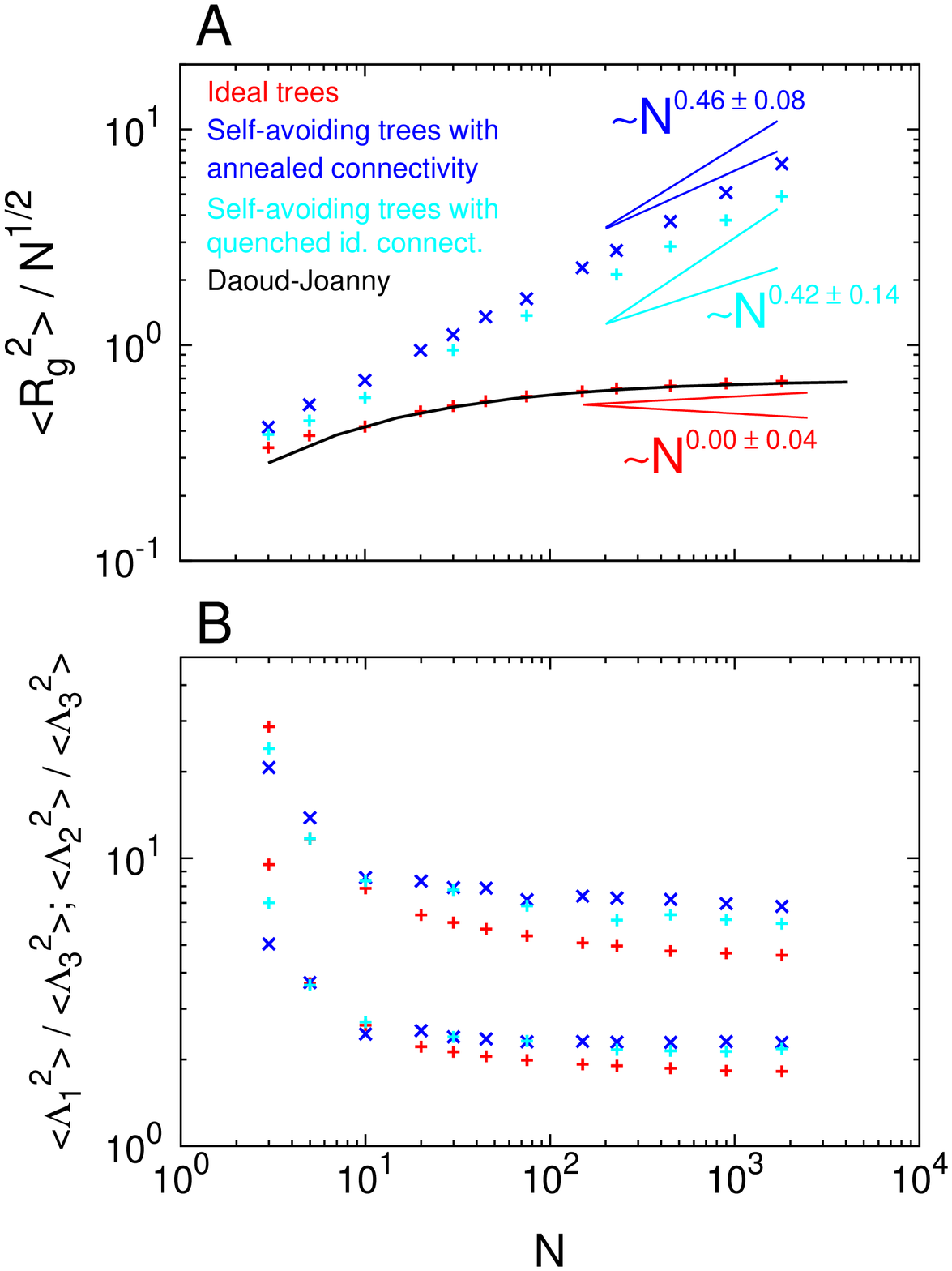}
\end{center}
\caption{
\label{fig:Rg2AspectRatios}
Conformational statistics of trees.
(A)
Mean square gyration radius, $\langle R_g^2 (N) \rangle$.
Straight lines correspond to the large-$N$ behaviour $\langle R_g^2(N) \rangle \sim N^{2 \nu}$ with critical exponents $\nu$
given by the best estimates summarised in Table~\ref{tab:ExpSummary}.
The black solid line corresponds to the analytical expression for ideal trees, Eq.~(\ref{eq:DaoudJoanny}).
(B)
Tree average aspect ratios, $\langle \Lambda_1^2 \rangle / \langle \Lambda_3^2 \rangle$ and $\langle \Lambda_2^2 \rangle / \langle \Lambda_3^2 \rangle$.
}
\end{figure}

Finally,
we have measured the mean-square gyration radius, $\langle R_g^2(N) \rangle$ (see tabulated values in Table~S5) and the average shape of trees as a function of tree weight, $N$
(Fig.~\ref{fig:Rg2AspectRatios}, panels A and B respectively).
Estimated values of critical exponents $\nu$ (straight lines in Fig.~\ref{fig:Rg2AspectRatios}A) are summarized in Table~\ref{tab:ExpSummary}.
In Table~S6 
we give details about their derivation.

\section{Discussion}\label{sec:Discussion}

\begin{table*}
\begin{tabular}{|c|c|c|c|c|c|c|c|}
\multicolumn{8}{c}{}\\
\hline
& & \multicolumn{2}{c|}{Ideal trees} & \multicolumn{2}{c|}{Self-avoiding trees,} & \multicolumn{2}{c|}{Self-avoiding trees,}\\
& & \multicolumn{2}{c|}{} & \multicolumn{2}{c|}{annealed connectivity} & \multicolumn{2}{c|}{quenched ideal connect.}\\
\hline
& {Relation to} & {Theory} & {Simul.} & {Theory} & {Simul.} & {Theory} & {Simul.}\\
& {other exponents} & & & & & &\\
\hline
& & & & & & & \\
{$\rho$} & -- & {$\frac{1}{2} = 0.5$} & {$0.49 \pm 0.04$} & {$\frac{9}{13} \approx 0.692$} & {$0.64 \pm 0.02$} & -- & --\\
& & & & & & & \\
{$\epsilon$} & {$=\rho$} & {$\frac{1}{2} = 0.5$} & {$0.536 \pm 0.007$} & {$\frac{9}{13} \approx 0.692$} & {$0.655 \pm 0.009$} & -- & --\\
& & & & & & & \\
{$\nu_{path}$} & -- & {$\frac{1}{2} = 0.5$} & {$0.509 \pm 0.008$} & {$\frac{7}{9} \approx 0.778$} & {$0.74 \pm 0.02$} & {$1$} & {$0.87 \pm 0.03$}\\
& & & & & & & \\
{$\nu$} & {$= \rho \, \nu_{path}$} & {$\frac{1}{4} = 0.25$} & {$0.25 \pm 0.02$} & {$\frac{7}{13} \approx 0.539$} & {$0.48 \pm 0.04$} & {$\frac{1}{2} = 0.5$} & {$0.46 \pm 0.07$}\\
& & & & & & & \\
{$\theta_{path}$} & -- & -- & {$-0.04 \pm 0.04$} & -- & {$1.3 \pm 0.1$} & -- & {$1.5 \pm 0.1$}\\
& & & & & & & \\
\hline
\end{tabular}
\caption{
\label{tab:ExpSummary}
Critical exponents for $3d$ ideal and self-avoiding lattice trees.
(a)
$\rho$, $\epsilon$, $\nu_{path}$ and $\nu$:
Comparison between predictions of Flory theory and numerical results.
(b)
$\theta_{path}$:
Numerical results are obtained by combining the estimated values of $\nu_{path}$
with the values of ``$\nu_{path} \theta_{path}$''
averaged over the ranges of $l$'s where this quantity shows a quasi-plateau for $N=1800$ 
(see black lines in the inset of the r.h.s panels of Fig.~\ref{fig:PathR2_ContactFreq}):
$\nu_{path}\theta_{path} = -0.02 \pm 0.02$ (ideal trees);
$\nu_{path}\theta_{path} = 0.96 \pm 0.07$ (self-avoiding trees with annealed connectivity);
$\nu_{path}\theta_{path} = 1.34 \pm 0.04$ (self-avoiding trees with quenched ideal connectivity).
Average values and corresponding error bars have been rounded to the first significant decimal digit.
}
\end{table*}

We have analyzed the behavior of self-avoiding lattice trees with annealed and quenched ideal branching structures
in terms of a small set of exponents defined in the Introduction and in Section~\ref{sec:observables}.
Our results are summarised in Table~\ref{tab:ExpSummary}.
In particular, our estimate $\nu=0.48 \pm 0.04$ of the asymptotic value is in excellent agreement with the exact value~\cite{ParisiSourlasPRL1981} $\nu=1/2$ for self-avoiding trees with annealed connectivity.
The rest of the other exponents are in good qualitative agreement with the predictions from Flory theory~\cite{IsaacsonLubensky,DaoudJoanny1981,GutinGrosberg93,GrosbergSoftMatter2014}.
In particular, and as reported in previous Monte Carlo simulations~\cite{CuiChenPRE1996},
our estimate $\nu=0.46 \pm 0.07$ for self-avoiding trees with quenched ideal connectivity is slightly smaller than the annealed case.
In general, annealed systems swell by a combination of modified branching and path stretching, with the latter effect being dominant: 
$\nu_{path}-\nu_{path}^{ideal} > (\rho-\rho^{ideal})$ (Table~\ref{tab:ExpSummary}).
Our simulations directly confirm, that trees with quenched ideal connectivity exhibit {\em less} overall swelling in good solvent than corresponding trees with annealed connectivity (Fig.~\ref{fig:Rg2AspectRatios}A),
even though they are {\em more strongly} stretched on the path level (Fig.~S3). 

Of course, the reader will notice that
there are deviations between observed exponents and those predicted by Flory theory (Table~\ref{tab:ExpSummary}).
This is not entirely surprising, as the approach is far from exact~\cite{DeGennesBook}.
This holds, of course, independently of the fact that we are dealing with trees. For the ``target'' of Flory theory, the exponent $\nu$, these deviations are surprisingly small: in $d=3$ the theory predicts for self avoiding walks~\cite{Flory1969} 
$\nu=3/5$ instead of the best theoretical estimate~\cite{LeGuillouZinnJustin} of $\nu=0.588$; similarly, in the case of annealed self-avoiding trees, Flory theory~\cite{GutinGrosberg93} predicts $\nu\approx0.54$ instead of $\nu=0.5$ \cite{ParisiSourlasPRL1981}. 
For other quantities, Flory theory is even qualitatively wrong. 
A particularly interesting case are the average contact probability $\langle p_c(l) \rangle$ between nodes at path distance $l$.
As shown in Fig.~\ref{fig:PathR2_ContactFreq} and Table~\ref{tab:ExpSummary},
$\langle p_c(l) \rangle$'s for interacting trees deviate consistently from the na\"ive mean-field estimate of $l^{-3\nu_{path}}$.
This is yet another illustration of the subtle cancellation of errors in Flory arguments, which are built on the mean-field estimates of contact probabilities~\cite{DeGennesBook}.

\section{Summary and Conclusion}\label{sec:concls}

In the present article, we have reconsidered the conformational statistics of lattice trees with volume interactions~\cite{IsaacsonLubensky,DaoudJoanny1981}. 
In particular, we have carried out computer simulations for two classes of three-dimensional systems,
namely self-avoiding lattice trees with annealed and quenched {\it ideal} connectivity (Sections~\ref{sec:AmoebaAlgo} and~\ref{sec:MDmethods}, respectively).
The well understood case of ideal, non-interacting lattice trees~\cite{ZimmStockmayer49,DeGennes1968} served as useful references.
In all cases, we have performed a detailed analysis of the observed connectivities and of the conformational statistics (Sections~\ref{sec:burning} and~\ref{sec:ExtractExps}).
In particular, we have determined (Section~\ref{sec:results}) values of suitable exponents describing the scaling behavior of:
the average branch weight, $\langle N_{br} (N) \rangle \sim N^{\epsilon}$,
the average path length, $\langle L(N) \rangle \sim N^{\rho}$,
the tree and branch gyration radii, $\langle R_g^2(N) \rangle \sim N^{2\nu}$
and, addressed here for the first time, the mean-square path extension, $\langle R^2(l) \rangle \sim l^{2\nu_{path}}$, and the path mean contact probability $\langle p_c(l) \rangle \sim l^{-\nu_{path}(d+\theta_{path})}$.

The applicability of Flory theory~\cite{IsaacsonLubensky,DaoudJoanny1981,GutinGrosberg93,GrosbergSoftMatter2014} was not a foregone conclusion. In the case of linear chains, the approach is notorious (and appreciated) for the nearly perfect cancellation of large errors in the estimation of {\em both} terms in Eq.~(\ref{eq:fFlory}) \cite{DeGennesBook,DesCloizeauxBook}. This delicate balance might well have been destroyed for trees, where the Flory energy needs to be simultaneously minimized with respect to $L$ and $R$.
A strong point is the ability~\cite{GutinGrosberg93} of the approach to treat trees with annealed and with randomly quenched connectivity within the same formalism.
To reduce the number of unfavorable contacts, the latter can only swell in a manner analogous to linear chains~\cite{DeGennesBook} with $\nu_{path}> 1/2$,
while the former have the additional option of increasing their overall size by adjusting the branching statistics, $\rho>1/2$.
Our results confirm that annealed systems swell by the predicted~\cite{GutinGrosberg93} combination of modified branching and path stretching, and that 
trees with quenched ideal connectivity exhibit {\em less} overall swelling in good solvent than corresponding trees with annealed connectivity, even though they are {\em more strongly} stretched on the path level.

Our study also reveals some, not entirely unexpected, limitations of the available Flory theory for trees. 
First, the underlying uncontrolled approximations are manifest in (small) deviations of predicted from observed or exactly known values for the exponents defined in Eqs.~(\ref{eq:epsilon}) to (\ref{eq:nu_path}). 
Secondly, and in analogy to the case of linear chains, Flory theory fails more dramatically in predicting contact probabilities or entropies.
In a forthcoming publication~\cite{Rosa2016c}, we will try to extend the description of interacting trees beyond Flory theory by combining scaling arguments with the wealth of information available from the analysis of  the computationally accessible distribution functions for the quantities, whose {\em mean} behaviour we have explored above.

{\it Acknowledgements} --
AR acknowledges grant PRIN 2010HXAW77 (Ministry of Education, Italy).
RE is grateful for the hospitality of the Kavli Institute for Theoretical Physics (Santa Barbara, USA)
and support through the National Science Foundation under Grant No. NSF PHY11-25915 during his visit in 2011, which motivated the present work. In particular, we have benefitted from long and stimulating exchanges with M. Rubinstein and A. Yu. Grosberg on the theoretical background.
This work was only possible thanks to generous grants of computer time by PSMN (ENS-Lyon) and P2CHPD (UCB Lyon 1), in part through the equip@meso facilities of the FLMSN.


\clearpage

\setcounter{section}{0}
\setcounter{figure}{0}
\setcounter{table}{0}
\setcounter{equation}{0}

\renewcommand{\figurename}{Fig. S}
\renewcommand{\tablename}{Table S}

{\large \bf Supplementary Data}

\tableofcontents

\clearpage

\section{Supplementary Figures}

\begin{figure*}
\includegraphics[width=3.4in]{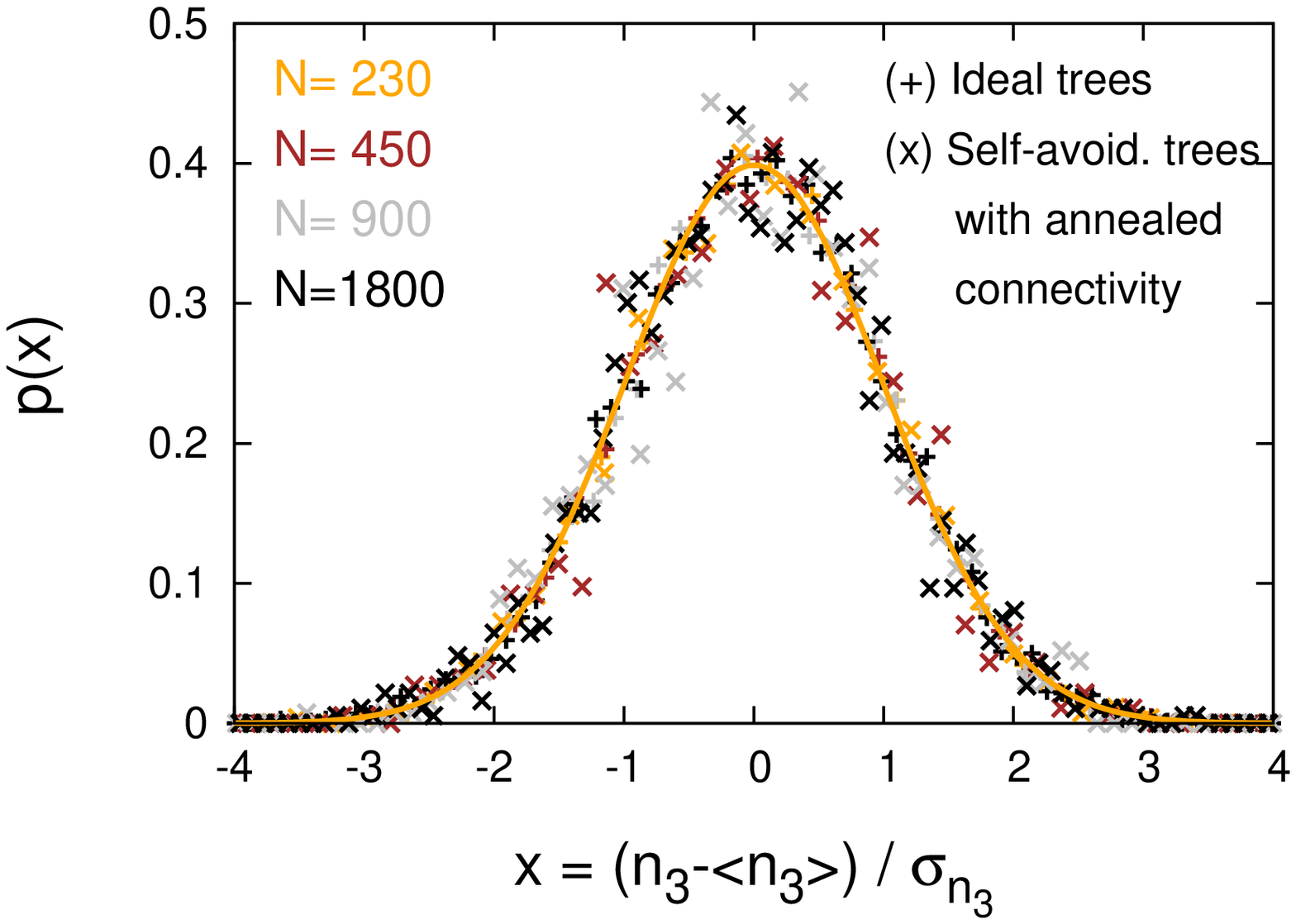}
\caption{
\label{fig:3fDistrs}
Branching statistics.
Distribution functions for the number of branching points $n_3$ (symbols) follow the Gaussian distribution (orange solid line).
Corresponding variances $\sigma_{n_3}^2$ increase linearly with $N$ as:
$\sigma_{n_3}^2 / N = 0.0410 \pm 0.0002$ (ideal trees); 
$\sigma_{n_3}^2 / N = 0.0633 \pm 0.0007$ (self-avoiding trees with annealed connectivity).
}
\end{figure*}

\begin{figure*}
\includegraphics[width=\textwidth]{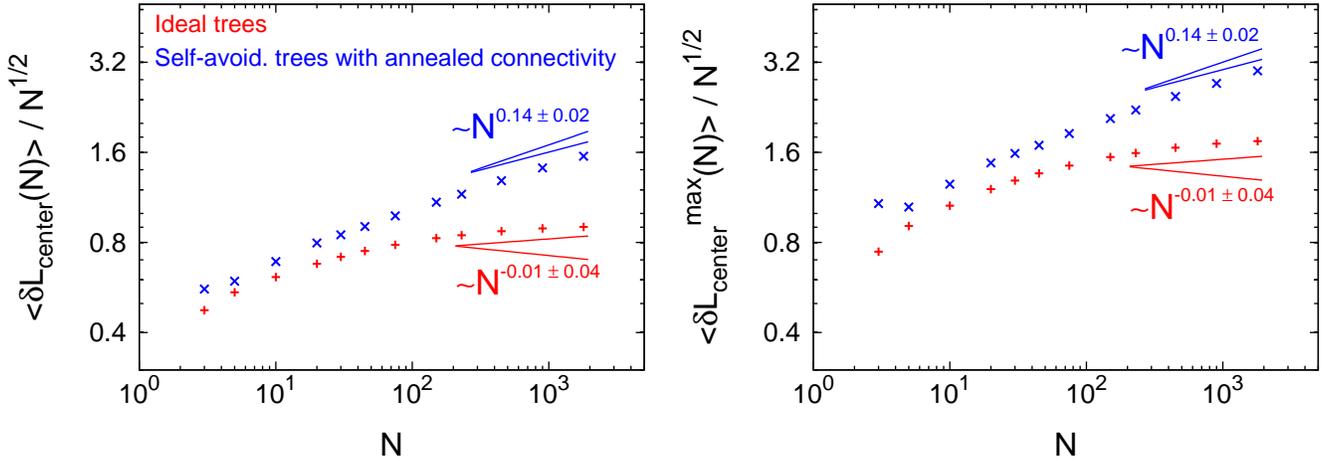}
\caption{
\label{fig:PathLengthStat_dLcenterdLcenterMax}
Path length statistics.
(Left)
Mean contour distance of nodes from the central node, $\langle \delta L_{center}(N) \rangle$.
(Right)
Mean longest contour distance of nodes from the central node, $\langle \delta L_{center}^{max}(N) \rangle$.
Straight lines correspond to the large-$N$ behaviour
$\langle \delta L_{center}(N) \rangle \sim \langle \delta L_{center}^{max}(N) \rangle \sim N^{\rho}$
with critical exponents $\rho$ given by the best estimates summarised in Table~1 main paper. 
}
\end{figure*}

\begin{figure*}
\includegraphics[width=\textwidth]{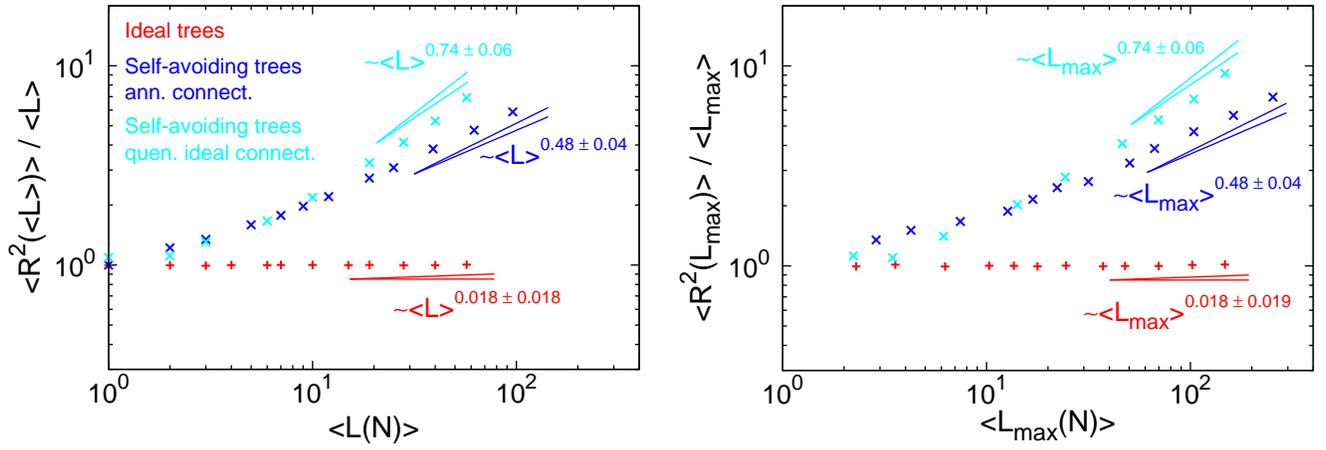}
\caption{
\label{fig:R2MaxLmax}
Conformational statistics of linear paths.
(Left)
Mean-square end-to-end distance, $\langle R^2(\langle L \rangle) \rangle$, of paths of length $l=\langle L(N) \rangle$.
(Right)
Mean-square end-to-end distance, $\langle R^2(L_{max}) \rangle$, of the longest paths.
Straight lines correspond to the large-$\langle L(N) \rangle$ or large-$\langle L_{max}(N) \rangle$ behaviours:
$\langle R^2(\langle L \rangle) \rangle \sim \langle L(N) \rangle^{2 \nu_{path}}$ or
$\langle R^2(L_{max}) \rangle \sim \langle L_{max}(N) \rangle^{2 \nu_{path}}$.
Critical exponents $\nu_{path}$ are given by the best estimates reported in Table~1 main paper. 
}
\end{figure*}

\clearpage

\section{Supplementary Tables}

\begin{table*}
\begin{tabular}{|c|c|c|c|c|c|}
\hline
\multicolumn{2}{|c}{} & \multicolumn{2}{|c|}{Ideal trees} & \multicolumn{2}{|c|}{Self-avoid. trees, annealed connectivity}\\
\hline
{$N$} & {$\tau_{MC}$} & {$R$} & {$\tau_{MC} / \tau_{corr}$} & {$R$} & {$\tau_{MC} / \tau_{corr}$}\\
\hline
{3} & {$1 \cdot 10^4$} & {$16000$} & {$\approx 1500$} & {$ 100$} & {$\approx 1650$}\\
{5} & {$1 \cdot 10^4$} & {$ 6400$} & {$\approx  400$} & {$ 100$} & {$\approx  625$}\\
{10} & {$1 \cdot 10^4$} & {$ 3200$} & {$\approx  100$} & {$ 100$} & {$\approx  155$}\\
{20} & {$1 \cdot 10^4$} & {$25000$} & {$\approx   25$} & {$1000$} & {$\approx   26$}\\
{30} & {$2 \cdot 10^4$} & {$25600$} & {$\approx   20$} & {$1000$} & {$\approx   26$}\\
{45} & {$5 \cdot 10^4$} & {$25600$} & {$\approx   20$} & {$1000$} & {$\approx   20$}\\
{75} & {$2 \cdot 10^5$} & {$25600$} & {$\approx   20$} & {$1000$} & {$\approx   20$}\\
{150} & {$1 \cdot 10^6$} & {$25600$} & {$\approx   13$} & {$1000$} & {$\approx   20$}\\
{230} & {$2 \cdot 10^6$} & {$25600$} & {$\approx   13$} & {$1000$} & {$\approx   17$}\\
{450} & {$1 \cdot 10^7$} & {$25600$} & {$\approx   13$} & {$1000$} & {$\approx   14$}\\
{900} & {$4 \cdot 10^7$} & {$12800$} & {$\approx   13$} & {$1000$} & {$\approx   11$}\\
{1800} & {$2 \cdot 10^8$} & {$ 6400$} & {$\approx   13$} & {$2000$} & {$\approx    9$}\\
\hline
\end{tabular}\\
\caption{
\label{tab:MCruns}
Details of Monte Carlo (MC) simulations for trees of mass $N$.
$\tau_{MC}$: MC steps per single run.
$R$: total number of independent runs.
$\tau_{MC} / \tau_{corr}$: total number of uncorrelated MC configurations per MC run.
$\tau_{corr}$ is the correlation time estimated {\it via} $g_3(\tau_{corr}) \approx \langle R_g^2\rangle$ (see also Fig.~1, main paper) 
where:
$g_3(\tau)$ is the mean-square displacement of the tree center of mass
as a function of MC steps 
and $\langle R_g^2\rangle$ is the tree mean-square gyration radius.
}
\end{table*}

\begin{table*}
\begin{tabular}{|c|c|c|c|c|c|c|}
\hline
$N$ & $R$ & $\tau_{MD} [\times 10^5]$ & $\tau_{MD} / \tau_{corr}$ & $\left \langle R_g^2 \right \rangle$ & $\tau_{MD} / \tau_{corr}$ & $\left \langle R_g^2 \right \rangle$ \\
\hline
\multicolumn{7}{c}{ }\\
\hline
\multicolumn{3}{|c}{} & \multicolumn{2}{|c|}{Self-avoid. trees, annealed connect.} & \multicolumn{2}{|c|}{Self-avoid. trees, quenched {\it ideal} connect.}\\
\hline
   3 & $100$ & $ 12$ & $\approx 5 \cdot 10^4$ & $  0.729 \pm 0.004$ & $\approx 5 \cdot 10^4$ & $  0.666 \pm 0.004$\\
   5 & $100$ & $ 12$ & $\approx 2 \cdot 10^4$ & $  1.093 \pm 0.009$ & $\approx 2 \cdot 10^4$ & $  0.997 \pm 0.007$\\
  10 & $100$ & $ 12$ & $\approx 7 \cdot 10^3$ & $  2.017 \pm 0.018$ & $\approx 7 \cdot 10^3$ & $  1.806 \pm 0.012$\\
  30 & $100$ & $ 12$ & $\approx 8 \cdot 10^2$ & $  5.948 \pm 0.057$ & $\approx 8 \cdot 10^2$ & $  5.193 \pm 0.054$\\
  75 & $100$ & $ 12$ & $\approx 10^3$ & $ 14.255 \pm 0.189$ & $\approx 10^3$ & $ 11.880 \pm 0.137$\\
 230 & $100$ & $ 12$ & $\approx 143$ & $ 41.847 \pm 0.550$ & $\approx 180$ & $ 32.163 \pm 0.478$\\
 450 & $100$ & $ 12$ & $\approx 39$ & $ 80.174 \pm 1.244$ & $\approx 54$ & $ 60.801 \pm 0.866$\\
 900 & $100$ & $ 12$ & $\approx 9$ & $157.470 \pm 2.837$ & $\approx 13$ & $113.660 \pm 1.741$\\
1800 & $100$ & $ 12$ & $\approx 3$ & $305.323 \pm 7.277$ & $\approx 4$ & $207.626 \pm 4.742$\\
\hline
\end{tabular}
\caption{
\label{tab:MDruns}
Details of Molecular Dynamics (MD) simulations for self-avoiding trees of mass $N$.
Quantities are defined as in Table~S\ref{tab:MCruns}
with the exception for $\tau_{MD}$ which represents the total length of a MD trajectory in standard time units, $\tau_{LJ}$~\cite{lammps}.
In analogy with Table~S\ref{tab:MCruns}, $\tau_{corr}$ is defined {\it via} $g_3(\tau_{corr}) \approx \langle R_g^2 \rangle$
(see also Fig.~2, main paper). 
}
\end{table*}

\begin{table*}
\begin{tabular}{|c|c|c|c|c|c|}
\hline
{$N$} & {$\langle n_3\rangle$} & {$\langle \delta L_{center} \rangle$} & {$\langle \delta L_{center}^{max} \rangle$} & {$\langle L \rangle$} & {$\langle N_{br} \rangle $}\\
\hline
\multicolumn{6}{c}{ }\\
\hline
\multicolumn{6}{|c|}{Ideal trees}\\
\hline
{3} & {$  0.709 \pm 0.003$} & {$ 0.823 \pm 0.001$} & {$ 1.291 \pm 0.003$} & {$ 1.161 \pm 0.001$} & {$ 0.097 \pm 0.001$}\\
{5} & {$  1.450 \pm 0.015$} & {$ 1.220 \pm 0.001$} & {$ 2.030 \pm 0.002$} & {$ 1.692 \pm 0.001$} & {$ 0.464 \pm 0.001$}\\
{10} & {$  3.320 \pm 0.012$} & {$ 1.938 \pm 0.003$} & {$ 3.357 \pm 0.009$} & {$ 2.769 \pm 0.003$} & {$ 1.098 \pm 0.003$}\\
{20} & {$  7.305 \pm 0.006$} & {$ 3.036 \pm 0.002$} & {$ 5.393 \pm 0.004$} & {$ 4.399 \pm 0.002$} & {$ 2.131 \pm 0.002$}\\
{30} & {$ 11.256 \pm 0.007$} & {$ 3.923 \pm 0.003$} & {$ 7.054 \pm 0.005$} & {$ 5.715 \pm 0.003$} & {$ 2.971 \pm 0.003$}\\
{45} & {$ 17.202 \pm 0.008$} & {$ 5.025 \pm 0.004$} & {$ 9.137 \pm 0.007$} & {$ 7.358 \pm 0.004$} & {$ 4.017 \pm 0.004$}\\
{75} & {$ 29.120 \pm 0.012$} & {$ 6.800 \pm 0.006$} & {$12.512 \pm 0.011$} & {$10.008 \pm 0.006$} & {$ 5.706 \pm 0.006$}\\
{150} & {$ 58.874 \pm 0.017$} & {$10.124 \pm 0.010$} & {$18.897 \pm 0.021$} & {$14.974 \pm 0.010$} & {$ 8.874 \pm 0.011$}\\
{230} & {$ 90.614 \pm 0.019$} & {$12.834 \pm 0.012$} & {$24.137 \pm 0.023$} & {$19.027 \pm 0.013$} & {$11.453 \pm 0.011$}\\
{450} & {$177.882 \pm 0.026$} & {$18.510 \pm 0.018$} & {$35.180 \pm 0.037$} & {$27.526 \pm 0.020$} & {$16.879 \pm 0.019$}\\
{900} & {$356.425 \pm 0.051$} & {$26.740 \pm 0.038$} & {$51.330 \pm 0.069$} & {$39.850 \pm 0.043$} & {$24.724 \pm 0.039$}\\
{1800} & {$713.491 \pm 0.101$} & {$38.243 \pm 0.079$} & {$74.043 \pm 0.129$} & {$57.174 \pm 0.090$} & {$35.723 \pm 0.073$}\\
\hline
\multicolumn{6}{c}{ }\\
\hline
\multicolumn{6}{|c|}{Self-avoiding trees, annealed connectivity}\\
\hline
{3} & {$  0.130 \pm 0.034$} & {$ 0.968 \pm 0.008$} & {$  1.870 \pm 0.034$} & {$ 1.234 \pm 0.004$} & {$ 0.290 \pm 0.011$}\\
{5} & {$  0.740 \pm 0.061$} & {$ 1.328 \pm 0.014$} & {$  2.350 \pm 0.048$} & {$ 1.805 \pm 0.011$} & {$ 0.594 \pm 0.017$}\\
{10} & {$  2.050 \pm 0.091$} & {$ 2.185 \pm 0.023$} & {$  3.960 \pm 0.055$} & {$ 3.027 \pm 0.022$} & {$ 1.351 \pm 0.024$}\\
{20} & {$  4.652 \pm 0.036$} & {$ 3.565 \pm 0.014$} & {$  6.599 \pm 0.030$} & {$ 5.023 \pm 0.014$} & {$ 2.664 \pm 0.015$}\\
{30} & {$  7.400 \pm 0.046$} & {$ 4.649 \pm 0.019$} & {$  8.692 \pm 0.037$} & {$ 6.611 \pm 0.019$} & {$ 3.684 \pm 0.019$}\\
{45} & {$ 11.472 \pm 0.051$} & {$ 6.067 \pm 0.027$} & {$ 11.334 \pm 0.052$} & {$ 8.668 \pm 0.028$} & {$ 5.039 \pm 0.027$}\\
{75} & {$ 19.566 \pm 0.070$} & {$ 8.499 \pm 0.038$} & {$ 16.026 \pm 0.072$} & {$12.201 \pm 0.041$} & {$ 7.363 \pm 0.038$}\\
{150} & {$ 39.905 \pm 0.096$} & {$13.351 \pm 0.064$} & {$ 25.417 \pm 0.122$} & {$19.273 \pm 0.070$} & {$12.053 \pm 0.064$}\\
{230} & {$ 61.375 \pm 0.121$} & {$17.597 \pm 0.084$} & {$ 33.604 \pm 0.155$} & {$25.431 \pm 0.092$} & {$16.109 \pm 0.084$}\\
{450} & {$121.067 \pm 0.173$} & {$27.301 \pm 0.128$} & {$ 52.147 \pm 0.237$} & {$39.457 \pm 0.143$} & {$25.491 \pm 0.129$}\\
{900} & {$242.455 \pm 0.234$} & {$42.588 \pm 0.195$} & {$ 81.579 \pm 0.355$} & {$61.577 \pm 0.217$} & {$40.182 \pm 0.194$}\\
{1800} & {$485.551 \pm 0.335$} & {$65.865 \pm 0.219$} & {$127.182 \pm 0.588$} & {$95.730 \pm 0.246$} & {$62.711 \pm 0.218$}\\
\hline
\end{tabular}
\caption{
\label{tab:ConnectivityDataIdealTrees}
Connectivity and branching statistics for trees of mass $N$.
$\langle n_3 \rangle$, average number of three-functional nodes.
$\langle \delta L_{center} \rangle$, average path distance from the central node.
$\langle \delta L_{center}^{max} \rangle$, average longest path distance from the central node.
$\langle L \rangle$, average path distance between nodes.
$\langle N_{br} \rangle $, average branch weight. 
}
\end{table*}

\begin{table*}
\begin{tabular}{|c|c|c|c|c|}
\hline
{} & $\langle \delta L_{center} \rangle \sim N^{\rho}$ & $\langle \delta L_{center}^{max} \rangle\sim N^{\rho}$ & $\langle L \rangle \sim N^{\rho}$ & $\langle N_{br} \rangle \sim N^{\epsilon}$\\
\hline
\multicolumn{5}{c}{}\\
\hline
\multicolumn{5}{|c|}{Ideal trees}\\
\hline
$\Delta$ & $0.239 \pm 0.045$ & $0.280 \pm 0.044$ & $0.271 \pm 0.028$ & $0.706 \pm 0.032$\\
${\tilde \chi}^2$ & $1.672$ & $1.073$ & $2.355$ & $2.706$\\
${\mathcal Q}$ & $0.123$ & $0.376$ & $0.028$ & $0.013$\\
& $\rho = 0.446 \pm 0.013$ & $\rho = 0.467 \pm 0.010$ & $\rho = 0.457 \pm 0.007$ & $\epsilon = 0.529 \pm 0.002$\\
\hline
$\Delta$ & $0$ & $0$ & $0$ & $0$\\
${\tilde \chi}^2$ & $7.424$ & $11.399$ & $10.476$ & $12.172$\\
${\mathcal Q}$ & $0.006$ & $0.001$ & $0.001$ & $0.001$\\
& $\rho = 0.525 \pm 0.002$ & $\rho = 0.538 \pm 0.001$ & $\rho = 0.529 \pm 0.001$ & $\epsilon = 0.543 \pm 0.002$\\
\hline
& \multicolumn{3}{c|}{$\mathbf{\rho = 0.494 \pm 0.013 \pm 0.038}$} & $\mathbf{\epsilon = 0.536 \pm 0.002 \pm 0.007}$\\
\hline
\multicolumn{5}{c}{}\\
\hline
\multicolumn{5}{|c|}{Self-avoiding trees, annealed connectivity}\\
\hline
$\Delta$ & $0.417 \pm 0.137$ & $0.534 \pm 0.163$ & $0.741 \pm 0.129$ & $0.938 \pm 0.240$\\
${\tilde \chi}^2$ & $1.156$ & $1.077$ & $1.156$ & $1.043$\\
${\mathcal Q}$ & $0.327$ & $0.374$ & $0.327$ & $0.395$\\
& $\rho = 0.632 \pm 0.017$ & $\rho = 0.639 \pm 0.012$ & $0.640 \pm 0.004$ & $\epsilon = 0.661 \pm 0.006$\\
\hline
$\Delta$ & $0$ & $0$ & $0$ & $0$\\
${\tilde \chi}^2$ & $0.636$ & $0.101$ & $0.209$ & $0.762$\\
${\mathcal Q}$ & $0.425$ & $0.750$ & $0.647$ & $0.383$\\
& $\rho = 0.635 \pm 0.004$ & $\rho = 0.643 \pm 0.005$ & $\rho = 0.639 \pm 0.003$ & $\epsilon = 0.649 \pm 0.004$\\
\hline
& \multicolumn{3}{c|}{$\mathbf{\rho = 0.638 \pm 0.017 \pm 0.004}$} & $\mathbf{\epsilon = 0.655 \pm 0.006 \pm 0.006}$\\
\hline
\end{tabular}
\caption{
\label{tab:FitsCritExpsTreeStructure}
Critical exponents $\rho$ and $\epsilon$,
describing path length ($\langle \delta L_{center} \rangle \sim \langle \delta L_{center}^{max} \rangle \sim \langle L \rangle \sim N^{\rho}$)
and branching statistics ($\langle N_{br}(N) \rangle \sim N^{\epsilon}$), respectively.
Single estimates were obtained from best fits of
data with $N \geq 450$ 
to simple power-law behaviour ($\Delta=0$) and
data with $N \geq 10$ to power-law behaviour with a correction-to-scaling term ($\Delta > 0$).
Combined together, they provide the final estimates shown in boldface with
uncertainties reported as ``$\pm \mbox{ statistical error} \pm \mbox{systematic error}$''.
For more details on the fitting procedure, see section~3.4 main paper. 
}
\end{table*}

\begin{table*}
\begin{tabular}{|c|c|c|c|c|}
\hline
{$N$} & {$\langle R^2 (\langle L \rangle) \rangle$} & {$\langle L_{max} \rangle$} & {$\langle R^2 (L_{max}) \rangle$} & {$\left \langle R_g^2 \right \rangle $} \\
\hline
\multicolumn{5}{c}{ }\\
\hline
\multicolumn{5}{|c|}{Ideal trees}\\
\hline
{3} & {$1.000 \pm 0.000$} &  {$2.291 \pm 0.004$} & {$2.276 \pm 0.012$} & {$ 0.578 \pm 0.001$} \\
{5} & {$1.994 \pm 0.006$} & {$3.566 \pm 0.007$} & {$3.617 \pm 0.033$} & {$ 0.851 \pm 0.010$} \\
{10} & {$2.978 \pm 0.015$} & {$6.257 \pm 0.014$} & {$6.207 \pm 0.085$} & {$ 1.323 \pm 0.028$} \\
{20} & {$3.998 \pm 0.006$} & {$10.287 \pm 0.009$} & {$10.305 \pm 0.052$} & {$ 2.199 \pm 0.005$} \\
{30} & {$5.998 \pm 0.011$} & {$13.615 \pm 0.011$} & {$13.604 \pm 0.069$} & {$ 2.857 \pm 0.006$} \\
{45} & {$7.018 \pm 0.012$} & {$17.774 \pm 0.015$} & {$17.668 \pm 0.090$} & {$ 3.688 \pm 0.008$} \\
{75} & {$10.012 \pm 0.018$} & {$24.526 \pm 0.022$} & {$24.587 \pm 0.126$} & {$ 5.006 \pm 0.012$} \\
{150} & {$14.974 \pm 0.027$} & {$37.291 \pm 0.035$} & {$37.094 \pm 0.192$} & {$ 7.468 \pm 0.016$} \\
{230} & {$19.034 \pm 0.035$} & {$47.774 \pm 0.045$} & {$47.664 \pm 0.250$} & {$ 9.528 \pm 0.021$} \\
{450} & {$27.931 \pm 0.052$} & {$69.862 \pm 0.067$} & {$69.747 \pm 0.362$} & {$13.718 \pm 0.028$} \\
{900} & {$39.923 \pm 0.104$} & {$102.155 \pm 0.139$} & {$103.217 \pm 0.764$} & {$19.936 \pm 0.063$} \\
{1800} & {$57.296 \pm 0.208$} & {$147.584 \pm 0.283$} & {$149.921 \pm 1.575$} & {$28.802 \pm 0.122$} \\
\hline
\multicolumn{5}{c}{ }\\
\hline
\multicolumn{5}{|c|}{Self-avoiding trees, annealed connectivity}\\
\hline
{3} & {$1.000 \pm 0.000$} & {$2.870 \pm 0.034$} & {$3.870 \pm 0.162$} & {$0.721 \pm 0.015$} \\
{5} & {$2.442 \pm 0.037$} & {$4.260 \pm 0.061$} & {$6.420 \pm 0.314$} & {$1.183 \pm 0.030$} \\
{10} & {$4.038 \pm 0.072$} & {$7.440 \pm 0.100$} & {$12.400 \pm 0.816$} & {$2.169 \pm 0.068$} \\
{20} & {$7.955 \pm 0.044$} & {$12.692 \pm 0.057$} & {$23.832 \pm 0.476$} & {$4.220 \pm 0.036$} \\
{30} & {$12.430 \pm 0.074$} & {$16.860 \pm 0.073$} & {$36.290 \pm 0.752$} & {$6.113 \pm 0.056$} \\
{45} & {$17.759 \pm 0.100$} & {$22.190 \pm 0.102$} & {$54.594 \pm 1.087$} & {$9.073 \pm 0.083$} \\
{75} & {$26.479 \pm 0.145$} & {$31.551 \pm 0.143$} & {$83.281 \pm 1.721$} & {$14.189 \pm 0.126$} \\
{150} & {$51.827 \pm 0.281$} & {$50.343 \pm 0.242$} & {$164.553 \pm 3.382$} & {$27.924 \pm 0.256$} \\
{230} & {$77.056 \pm 0.429$} & {$66.715 \pm 0.308$} & {$257.855 \pm 5.248$} & {$41.673 \pm 0.365$} \\
{450} & {$149.150 \pm 0.811$} & {$103.808 \pm 0.475$} & {$487.242 \pm 9.673$} & {$79.386 \pm 0.666$} \\
{900} & {$294.509 \pm 1.600$} & {$162.655 \pm 0.710$} & {$920.281 \pm 19.119$} & {$152.510 \pm 1.244$} \\
{1800} & {$563.185 \pm 2.027$} & {$254.028 \pm 0.836$} & {$1774.609 \pm 25.199$} & {$293.656 \pm 2.441$} \\
\hline
\multicolumn{5}{c}{ }\\
\hline
\multicolumn{5}{|c|}{Self-avoiding trees, quenched {\it ideal} connectivity}\\
\hline
{3} & {$1.091 \pm 0.007$} & {$2.220 \pm 0.013$} & {$2.491 \pm 0.056$} & {$0.666 \pm 0.004$} \\
{5} & {$2.229 \pm 0.021$} & {$3.460 \pm 0.018$} & {$3.802 \pm 0.086$} & {$0.997 \pm 0.007$} \\
{10} & {$3.927 \pm 0.031$} & {$6.150 \pm 0.023$} & {$8.645 \pm 0.179$} & {$1.806 \pm 0.012$} \\
{30} & {$9.996 \pm 0.069$} & {$14.160 \pm 0.059$} & {$28.642 \pm 0.584$} & {$5.193 \pm 0.054$} \\
{75} & {$21.855 \pm 0.124$} & {$24.290 \pm 0.109$} & {$67.453 \pm 1.383$} & {$11.880 \pm 0.137$} \\
{230} & {$61.987 \pm 0.299$} & {$46.340 \pm 0.224$} & {$189.496 \pm 3.603$} & {$32.163 \pm 0.478$} \\
{450} & {$115.740 \pm 0.564$} & {$69.610 \pm 0.298$} & {$373.776 \pm 6.824$} & {$60.801 \pm 0.866$} \\
{900} & {$211.451 \pm 0.966$} & {$103.890 \pm 0.494$} & {$709.591 \pm 14.394$} & {$113.660 \pm 1.741$} \\
{1800} & {$394.470 \pm 1.693$} & {$147.670 \pm 0.743$} & {$1354.818 \pm 25.224$} & {$207.626 \pm 4.742$} \\
\hline
\end{tabular}
\caption{
\label{tab:ConformationalDataIdealTrees}
Conformational statistics of lattice trees of mass $N$.
$\langle R^2 (\langle L \rangle) \rangle$, mean-square end-to-end distance of paths of length $l = \langle L(N) \rangle$.
$\langle L_{max} \rangle$, average length of the longest paths.
$\langle R^2 (L_{max}) \rangle$, mean-square end-to-end distance of the longest paths.
$\left \langle R_g^2 \right \rangle$, mean-square gyration radius.
}
\end{table*}

\begin{table*}
\begin{tabular}{|c|c|c|c|}
\hline
{} & $\langle R^2 (\langle L \rangle) \rangle \sim \langle L(N) \rangle^{2\nu_{path}}$ & $\langle R^2 (L_{max}) \rangle \sim \langle L_{max}(N) \rangle^{2\nu_{path}}$ & $\left \langle R_g^2 \right \rangle \sim N^{2\nu}$ \\
\hline
\multicolumn{4}{c}{ }\\
\hline
\multicolumn{4}{|c|}{Ideal trees}\\
\hline
$\Delta$ & -- & -- & $0.277 \pm 0.105$ \\
${\tilde \chi}^2$ & -- & -- & $1.770$ \\
${\mathcal Q}$ & -- & -- & $0.101$ \\
& -- & -- & $\nu = 0.229 \pm 0.013$ \\
\hline
$\Delta$ & $0$ & $0$ & $0$ \\
${\tilde \chi}^2$ & $0.999$ & $0.112$ & $0.562$ \\
${\mathcal Q}$ & $0.318$ & $0.738$ & $0.454$ \\
& $\nu_{path} = 0.504 \pm 0.003$ & $\nu_{path} = 0.513 \pm 0.007$ & $\nu = 0.268 \pm 0.002$ \\
\hline
& \multicolumn{2}{|c|}{$\mathbf{\nu_{path} = 0.509 \pm 007 \pm 0.005}$} & $\mathbf{\nu = 0.249 \pm 0.013 \pm 0.020}$ \\
\hline
\multicolumn{4}{c}{ }\\
\hline
\multicolumn{4}{|c|}{Self-avoiding trees, annealed connectivity}\\
\hline
$\Delta$ & $1.018 \pm 0.213$ & -- & $0.294 \pm 0.273$ \\
${\tilde \chi}^2$ & $0.922$ & -- & $1.379$ \\
${\mathcal Q}$ & $0.478$ & -- & $0.219$ \\
& $\nu_{path} = 0.748 \pm 0.006$ & -- & $\nu = 0.486 \pm 0.036$ \\
\hline
$\Delta$ & $0$ & $0$ & $0$ \\
${\tilde \chi}^2$ & $0.291$ & $0.279$ & $0.013$ \\
${\mathcal Q}$ & $0.589$ & $0.597$ & $0.910$ \\
& $\nu_{path} = 0.738 \pm 0.004$ & $\nu_{path} = 0.723 \pm 0.014$ & $\nu = 0.472 \pm 0.004$ \\
\hline
& \multicolumn{2}{|c|}{$\mathbf{\nu_{path} = 0.736 \pm 0.014 \pm 0.010}$} & $\mathbf{\nu = 0.479 \pm 0.036 \pm 0.007}$\\
\hline
\multicolumn{4}{c}{ }\\
\hline
\multicolumn{4}{|c|}{Self-avoiding trees, quenched {\it ideal} connectivity}\\
\hline
$\Delta$ & $0.691 \pm 0.258$ & -- & -- \\
${\tilde \chi}^2$ & $5.899$ & -- & -- \\
${\mathcal Q}$ & $0.0005$ & -- & -- \\
& $0.899 \pm 0.021$ & -- & -- \\
\hline
$\Delta$ & $0$ & $0$ & $0$ \\
${\tilde \chi}^2$ & $5.065$ & $3.415$ & $0.784$ \\
${\mathcal Q}$ & $0.024$ & $0.065$ & $0.376$ \\
& $\nu_{path} = 0.863 \pm 0.005$ & $\nu_{path} = 0.855 \pm 0.017$ & $\nu = 0.463 \pm 0.008$ \\
\hline
& \multicolumn{2}{|c|}{$\mathbf{\nu_{path} = 0.872 \pm 0.021 \pm 0.019}$} & $\mathbf{\nu = 0.463 \pm (0.072) \pm (0.014)}$ \\
\hline
\end{tabular}
\caption{
\label{tab:FitsCritExpsSpatialStructure}
Critical exponents $\nu_{path}$ and $\nu$ 
describing, respectively, the scaling behaviors of
$\langle R^2 (\langle L \rangle) \rangle \sim \langle L(N) \rangle^{2\nu_{path}}$ and $\langle R^2(L_{max}) \rangle \sim \langle L_{max}(N) \rangle^{2\nu_{path}}$,
and
$\langle R_g^2(N) \rangle \sim \langle N\rangle^{2\nu}$.
Single estimates for $\nu$ are the results of fitting the data to
three- ($\Delta > 0$ for data with $N \geq 10$) and
two-parameter ($\Delta = 0$ for data with $N \geq 450$)
functions, as described in section~3.4 main paper. 
In the case of self-avoiding trees with quenched ideal connectivity the three-parameter fits fail,
and corresponding statistical and systematic errors (in brackets) have been fixed on the same trends observed for annealed self-avoiding trees:
``error 2-parameter fit'' $\approx \frac{1}{9}$ ``error 3-parameter fit'' $\approx \frac{1}{2}$ ``systematic error''.
For the critical exponent $\nu_{path}$,
we have attempted similar analyses with two- and three-parameter fit functions:
$\log \langle R^2(\langle L \rangle) \rangle = a + 2\nu_{path} \log \langle L(N) \rangle$ and
$\log \langle R^2(\langle L \rangle) \rangle = a + b\langle L(N) \rangle^{-\Delta_0} + 2\nu_{path} \log \langle L(N) \rangle - b (\Delta-\Delta_0) \langle L(N) \rangle^{-\Delta_0} \log \langle L(N) \rangle$,
and analogous expressions with $\langle L_{max}(N) \rangle$.
When three-parameter fits fail, only the results from two-parameter fits are combined into our final estimates for $\nu_{path}$ (in boldface). 
}
\end{table*}

\end{document}